\documentclass[
    aps,
    prd,
    superscriptaddress,
    tightenlines,
    nofootinbib,
    floatfix,
    longbibliography,
    notitlepage,
    twocolumn,
    preprintnumbers
]{revtex4-2}

\usepackage[T1]{fontenc}
\usepackage[utf8]{inputenc}
\usepackage[UKenglish]{babel}
\usepackage{physics}
\usepackage[dvipsnames]{xcolor}
\usepackage{listings}
\usepackage[super]{nth}
\usepackage{mathtools}
\usepackage{nicefrac}
\usepackage{amsmath}
\usepackage{amssymb}
\usepackage{amsfonts}
\usepackage{dsfont}
\usepackage{upgreek}
\usepackage{hyperref}
\usepackage{booktabs}
\usepackage{comment}
\usepackage{float}
\usepackage{siunitx}
\usepackage{slashed}

\usepackage[caption=false,position=top,labelformat=parens]{subfig}

\usepackage{tikz}
    \usetikzlibrary{
        mindmap,
        shadows,
        arrows,
        matrix,
        arrows.meta,
        shapes,
        decorations,
        decorations.shapes,
        decorations.markings,
        decorations.pathmorphing
    }
    \tikzstyle{arrow} = [very thick,->,>=stealth]
    \tikzstyle{middleArrow} = [-, decoration={markings, mark=at position 0.5 with {\arrow{Stealth}} }, line width = 0.1em, postaction={decorate} ]
    \tikzstyle{snakeLine} = [-, decorate, decoration={snake,amplitude=0.2em,segment length=0.6em}, line width = 0.1em, ]
    \tikzstyle{latticeLine} = [-, line width=0.1em]
    \tikzstyle{latticeLink} = [-, decoration={markings, mark=at position 0.6 with {\arrow{Stealth}} }, line width = 0.1em, fzjred, postaction={decorate} ]

\usepackage{ulem}
\usepackage{placeins} 
\hypersetup{colorlinks, linkcolor = [rgb]{0,0.0,0.75}, citecolor = [rgb]{0,0.0,0.75}, urlcolor = [rgb]{0,0.0,0.75}}

\newcommand{\im}{\ensuremath{\mathrm{i}}}

\newcommand{\DLR}{\ensuremath{\overset{\leftrightarrow}{D}}}
\newcommand{\DL}{\ensuremath{\overset{\leftarrow}{D}}}
\newcommand{\DR}{\ensuremath{\overset{\rightarrow}{D}}}

\renewcommand{\O}{\ensuremath{\mathcal{O}}}

\newcommand{\MSbar}{\ensuremath{\overline{\mathrm{MS}}} }

\newcommand{\Ncfg}{\ensuremath{N_{\mathrm{cfg}}}}

\renewcommand{\bar}[1]{\ensuremath{\overline{#1}}}

\newcommand{\COpt}[3][]{\ensuremath{\mathrm{C}_{3\mathrm{pt}}^{#1}\left(#2,\,#3\right)}}

\newcommand{\Cpt}[1]{\ensuremath{}\mathrm{C}_{2\mathrm{pt}}\left(#1\right)}

\definecolor{fzjblue}{RGB}{2,61,107}
\definecolor{fzjlightblue}{RGB}{173,189,227}
\definecolor{fzjgray}{RGB}{235,235,235}
\definecolor{fzjred}{RGB}{235, 95, 115}
\definecolor{fzjgreen}{RGB}{185, 210, 95} 
\definecolor{fzjyellow}{RGB}{250, 235, 90}
\definecolor{fzjviolet}{RGB}{175, 130, 185}
\definecolor{fzjorange}{RGB}{250, 180, 90}
\definecolor{fzjblack}{RGB}{0,0,0}
\definecolor{fzjwhite}{RGB}{255,255,255}

\newcommand{\jsc}{
    J\"{u}lich Supercomputing Center \& Institute for Advanced Simulation \& Center for Advanced Simulation and Analytics (CASA),
    Forschungszentrum J\"{u}lich, 52428 J\"{u}lich, Germany\\
}
\newcommand{\bonn}{
    Helmholtz-Institut f\"{u}r Strahlen- und Kernphysik,
    Rheinische Friedrich-Wilhelms-Universit\"{a}t Bonn, 53115 Bonn, Germany
}
\newcommand{\mitctp}{
    Center for Theoretical Physics,  
    MIT,  Cambridge, MA 02139, USA
}
\newcommand{\suny}{
    Department of Physics and Astronomy, 
    Stony Brook University, Stony Brook, NY 11794, USA
}
\newcommand{\bnl}{
    RIKEN/BNL  Research  Center,  
    Brookhaven  National  Laboratory,  Upton, NY 11973, USA
}
\newcommand{\nmsu}{
    Department of Physics,
    New Mexico State University, Las Cruces, NM 88003, USA
}
\newcommand{\UA}{
    Department of Physics,
    University  of  Arizona,  Tucson,  AZ 85721, USA
}
\newcommand{\zppt}{
    Deutsches Elektronen-Synchrotron DESY,
    Platanenallee 6, 15738 Zeuthen, Germany
}
\newcommand{\uva}{
    Department of Physics,
    University of Virginia, Charlottesville, VA 22904, USA
}

\date{\today}

\begin{document}

\preprint{DESY-23-212}

\title{Moments of Nucleon Unpolarized, Polarized, and Transversity Parton Distribution Functions from Lattice QCD at the Physical Point}
\author{Marcel~Rodekamp}
   \affiliation{\jsc}
   \affiliation{\bonn}
\author{Michael~Engelhardt}
   \affiliation{\nmsu}
\author{Jeremy~R.~Green}
   \affiliation{\zppt}
\author{Stefan~Krieg}
   \affiliation{\jsc}
   \affiliation{\bonn}
\author{Simonetta~Liuti}
   \affiliation{\uva}
\author{Stefan~Meinel}
   \affiliation{\UA}
\author{John~W.~Negele}
   \affiliation{\mitctp}
\author{Andrew~Pochinsky}
    \affiliation{\mitctp}
\author{Sergey~Syritsyn}
    \affiliation{\suny}
    \affiliation{\bnl}
\begin{abstract}
The second Mellin moments $\expval{x}$ of the nucleon's unpolarized, polarized, and transversity parton distribution functions (PDFs) are computed.
Two lattice QCD ensembles at the physical pion mass are used:
these were generated using a tree-level Symanzik-improved gauge action
and 2+1 flavour tree-level improved Wilson Clover fermions coupling via 2-level HEX-smearing.
The moments are extracted from forward matrix elements of local leading twist operators.
We determine renomalization factors in RI-(S)MOM and match to $\bar{\mathrm{MS}}$ at scale $2\,\si{GeV}$.
Our findings show that operators that exhibit vanishing kinematics at zero momentum can have significantly reduced excited-state contamination.
The resulting polarized moment is used to quantify the longitudinal contribution to the quark spin-orbit correlation.
All our results agree within two sigma with previous lattice results.
\end{abstract}
 \maketitle

\section{Introduction}\label{sec-Introduction}
The distribution of the momentum and spin within a hadron is encoded by parton distribution functions (PDFs). 
Determining the PDFs is thus an indispensable ingredient to our understanding of the structure of hadrons~\cite{sterman_handbook_1995,lai_new_2010,ji_parton_2013}.
There have been various efforts of extracting the PDFs from global fits, for a recent summary see~\cite{li_extraction_2023}.
The Lattice QCD community has also achieved remarkable strides in the computation of PDFs over the recent years~\cite{linPartonDistributionsLattice2018,cichy2019guide,Constantinou:2022yye}.

In this study\footnote{Preliminary results were reported in \cite{rodekamp_moments_2023}.}, our focus centers on the evaluation of the second Mellin moment, denoted as $\expval{x}$~\cite{gockeler_lattice_1996,haglerHadronStructureLattice2010,linPartonDistributionsLattice2018,constantinouXdependenceHadronicParton2021},
of unpolarized, polarized, and transversity PDFs.
We achieve this through the examination of matrix elements of local twist-two operators~\cite{gockelerQuarkHelicityFlip2005,haglerHadronStructureLattice2010, alexandrou2016parton,harris_nucleon_2019,bali_nucleon_2019,alexandrou_moments_2020,mondal_moments_2020}.
This method does not require high momenta to suppress higher-twist contributions, as is needed in calculations that use non-local operators, for example the widely used quasi-PDF method~\cite{ji_parton_2013, cichy2019guide}.
One of our objectives is to gain insights into the contamination stemming from excited states for different matrix elements and constraining the resulting uncertainty.
To attain this objective, a comprehensive investigation of matrix elements at finite but modest momenta becomes imperative, as certain operators have nonvanishing matrix elements exclusively at nonzero momentum. 
Although the exploration of forward matrix elements of local operators at non-zero momentum is somewhat unconventional, references~\cite{brattNucleonStructureMixed2010,therqcdcollaborationNucleonAxialStructure2020,barcaPiMatrixElements2022} have previously ventured into this territory.

This paper is organized as follows. 
In section~\ref{sec-Method} we explain our analysis chain and discuss in detail 
which operators are considered.
The different steps of the analysis to extract the matrix elements are shown in section~\ref{subsec-Ratios}.
We continue the computation of moments in section~\ref{subsec-x} where they become renormalized and averaged over the different results. 
Further, our findings are put in relation to other Lattice QCD results and global fits.
In section~\ref{subsec-LS} we utilize the moment of the polarized PDF to compute the quark spin-orbit correlation.
Last, in section \ref{sec-Summary} we summarize our findings.
There are three appendices: appendix~\ref{apx-results-per-operator} shows extraction of the matrix element for each operator, which are summarized in appendix~\ref{apx:summaryPlots}. 
Finally, the calculation of renormalization factors is discussed in appendix~\ref{apx:npr}. 
 
\section{Method}\label{sec-Method}
Moments of PDFs can be obtained by calculating forward matrix elements of 
local leading-twist operators \cite{martinelli1988lattice,martinelli1989lattice,gockelerLatticeOperatorsMoments1996,alexandrou2016parton}
\begin{equation}
    \O^X \equiv \O^X_{\{ \alpha,\mu\}} =  \bar{q} \Gamma^X_{\{ \alpha} \DLR^{\phantom{X}}_{\mu \}} q. \label{eq-leading-twist-operator}
\end{equation}
Here, the symbol $X$ denotes either $V$, $A$, or $T$, corresponding to the vector, axial, or tensor channels, respectively, and in the tensor case $\Gamma_\alpha=\sigma_{\beta\gamma}$ so that $\alpha$ is a compound index. These channels are associated with unpolarized, polarized, or transversity PDFs.
Symmetrizing the indices and subtracting traces is indicated by braces, $\{\alpha,\mu\}$. 
We specifically focus on the isovector channel, which involves the difference between $\O^X$ for up and down quarks,
$\O^X(q=u)-\O^X(q=d)$, 
to avoid calculating disconnected diagrams. 
The left-right acting covariant derivative $\DLR = \nicefrac{1}{2}( \DR - \DL)$ is constructed on the Euclidean lattice using central finite differences between neighboring lattice points, connected by appropriate gauge links.

It is well understood that these forward matrix elements are proportional to the desired moment $\expval{x}$~\cite{gockelerQuarkHelicityFlip2005, haglerHadronStructureLattice2010, alexandrou2016parton}.
The matrix element is given by
\begin{equation}
\begin{aligned}
    \mel{N(p)}{\O^X_{\{\alpha,\mu\}}}{N(p)} 
                = \expval{x} \bar{u}_{N(p)} \Gamma^X_{\{ \alpha} \im p^{\phantom{X}}_{\mu \}} u_{N(p)}.\label{eq-matrix-element}
\end{aligned}
\end{equation}
In this equation, $p$ represents the 4-momentum of the nucleon.

In the continuum, the operators described in Equation~\eqref{eq-leading-twist-operator} form irreducible representations of the Lorentz group. 
However, in the context of Euclidean space, the Lorentz group is replaced by the orthogonal group~\cite{gockelerLatticeOperatorsMoments1996}. 
When we transition to the lattice, the orthogonal group further reduces to the hypercubic group $H(4)$. 
This reduced symmetry can lead to certain operators mixing with lower-dimensional ones. 
Fortunately, for the specific one-derivative operator studied, such mixing does not occur~\cite{gockeler2005lattice}.

Nevertheless, it is important to note that the Euclidean irreducible representations to which our operators belong are divided into multiple hypercubic irreducible representations. 
In our work, we adopt the common notation, where $\tau_a^{(b)}$ represents the $a^\mathrm{th}$ irreducible representation of dimension $b$. 
Each of these hypercubic irreducible representations necessitates a distinct renormalization factor. 
To keep renormalization diagonal, we construct operators with well-defined hypercubic irreducible representations, as suggested by Göckeler et al.\@~\cite{gockelerLatticeOperatorsMoments1996}.

In practical terms, this implies that for each $\tau_{a}^{(b)}$, we must compute the corresponding renormalization factor $Z_{\tau_{a}^{(b)}}$. 
This factor is subsequently applied to the matrix elements of an operator that transforms irreducibly under the given representation. 
As a result, we denote the renormalization factor for the operator $\O^X$ as $Z_{\O^X}$, equivalent to $Z_{\tau_{a}^{(b)}}$.

The matrix element described in Equation~\eqref{eq-matrix-element} can be determined on the lattice by considering the ratios of three-point and two-point correlation functions, as previously discussed in the literature, e.g.\@~\cite{gockelerQuarkHelicityFlip2005, alexandrou2016parton}. 
The two-point correlation function, denoted as 
\begin{equation}
\begin{gathered}
    \Cpt{\tau} = 
    \int \dd[3]{y} e^{-\mathrm{i} \vec{p}\vec{y}}  \Tr{\Gamma_\mathrm{pol} 
        \expval{
            \chi\left(\vec{y},\tau\right) \bar{\chi}\left(\vec{0},0\right) 
        }    
    }, 
\end{gathered}
\end{equation}
quantifies the correlation between a nucleon source and a nucleon sink separated by a time interval $\tau$.
Here we use\footnote{The same results can also be obtained using $P_+$ by itself as the spin projector in $C_\text{2pt}$.} $\Gamma_\mathrm{pol} = P_+\left[ 1-\mathrm{i}\gamma_1\gamma_2 \right]$ with $P_+ = \nicefrac{(1+\gamma_4)}{2}$ and a nucleon interpolating operator of the form
\begin{equation}
    \chi_\alpha = \epsilon_{abc} \left( \tilde{u}_{a}^T C \gamma_5 P_+ \tilde{d}_b \right) \tilde{u}_{c,\alpha}\label{eq-interpolating-field}
\end{equation}
with smeared quark fields $\tilde{q}$.

\begin{figure}
\resizebox{\linewidth}{!}{
\begin{tikzpicture}
    \node[fill=fzjviolet, draw=fzjviolet, circle, font=\large, text width = 2em ] at (0,0)  (NL)  {$\chi$};
    \node[fill=fzjviolet, draw=fzjviolet, circle, font=\large, text width = 2em ] at (10,0) (NR)  {$\chi$};

    \draw[fzjred  , line width = 2pt, line cap = round] (NL.north east) to[out=25,in=155] (NR.north west);
    \draw[fzjgreen, line width = 2pt, line cap = round] (NL.south east) to[out=335,in=205] (NR.south west);
    \draw[fzjblue , line width = 2pt, line cap = round] (NL.east) -- (NR.west);

    \node[fill=fzjlightblue, draw=fzjblue, ellipse, minimum height = 4em, font=\large] at (5,1.2) {$\O^X$};

    \draw[arrow] (-1,-2) -- (11,-2);
    \node[font=\large] at (0, -2.3) {0};
    \node[font=\large] at (5, -2.3) {$\tau$};
    \node[font=\large] at (10,-2.3) {$T$};

    \node[fill=fzjviolet, draw=fzjviolet, circle, font=\large ] at (0,0)  (NL) {$\chi$};
    \node[fill=fzjviolet, draw=fzjviolet, circle, font=\large ] at (10,0) (NR)  {$\chi$};
\end{tikzpicture}
}
\caption{
    Graphical representation of $\COpt[\O^X]{T}{\tau}$: a source 
    nucleon inserted at time $t=0$ and a sink nucleon removed at time $t=T$.
    A local leading twist operator~\eqref{eq-leading-twist-operator} is 
    inserted on a given time slice $\tau$.
    The nucleons on the lattice are represented by interpolating operators $\chi$~\eqref{eq-interpolating-field}
    while $\O^X$ is determined by finite differences connected with gauge links. 
}\label{fig-correlation}
\end{figure}
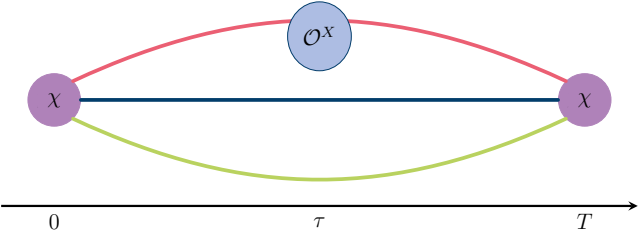

The three-point correlation function, denoted as 
\begin{multline}
    \COpt[\O^X]{T}{\tau} = 
        \int\dd[3]{y}\dd[3]{z} \Bigg[e^{-\mathrm{i} \vec{p}\,'\vec{y}} e^{\mathrm{i} \left(\vec{p}\,'-\vec{p}\right)\vec{z} }  \\
        \times \Tr{\Gamma_\mathrm{pol} 
            \expval{
                \chi\left(\vec{y},T\right)  \O^X\left(\vec{y},\tau\right) \bar{\chi}\left(\vec{0},0\right) 
            }
        }\Bigg],
\end{multline}
separates the source and sink nucleons by a time interval $T$ while incorporating the operator of interest, $\O^X$, at time $\tau$. 
From here on we let $\vec{p}\,' = \vec{p}$ as indicated in~\eqref{eq-matrix-element}.
A visual representation is given by Figure~\ref{fig-correlation}.
The matrix element is extracted in the limit where
\begin{align}
    \mathcal{M}
    &\equiv \lim_{T-\tau,\tau\to\infty} R(T,\tau) \\
    &\equiv \lim_{T-\tau,\tau\to\infty} \frac{ \COpt[\O^X]{T}{\tau}}{\Cpt{T}}.
    \label{eq-ratio-definition}
\end{align}
Once the matrix element is obtained, we can compute the moment by simply dividing the kinematic factor,
$
    \expval{x} K = \mathcal{M},
$
with
\begin{gather}
    K = 
    \frac{1}{2 E_N(p)} \\  \times\frac{
        \Tr{\Gamma_\mathrm{pol} \left( -\im \gamma_\mu p^\mu + m_N \right) \left[ a^{\alpha, \mu} \Gamma^X_{\alpha} p_{\mu} \right]\left( -\im \gamma_\mu p^\mu + m_N \right) }
    }{
        \Tr{\Gamma_\mathrm{pol} \left( -\im \gamma_\mu p^\mu + m_N \right)}
    } \nonumber,
\end{gather}
where the $a^{\alpha,\mu}\in\mathbb{R}$ are appropriate factors to express the symmetrization and removal of traces discussed above;

This analysis involves a spectral decomposition of the ratio, which allows us to isolate the matrix element of the ground state:
\begin{equation}
    R(T,\tau) = \mathcal{M} + \text{excited states}.\label{eq-1-state}
\end{equation}
To account for the influence of the first excited state, we expand the expression, obtaining the leading contribution from excited states
\begin{gather}
    \mathcal{M} 
    \frac{1 + R_1 e^{-\frac{T}{2}\Delta E } \cosh\left[ \left(\nicefrac{T}{2} - \tau \right) \Delta E\right] + R_2 e^{- T \Delta E} }{1 + R_3 e^{- T \Delta E}},\label{eq-2-state} 
\end{gather}
where $\Delta E$ represents the energy difference between the first excited state and the ground state ($\Delta E = E_1 - E_0$). 
In principle, one would aim to consider large values of $T$ and $\tau$ to approach the limit defined in Equation~\eqref{eq-ratio-definition}. 
However, it is important to note that as $T$ increases, so does the statistical noise.

The constants in the numerator, $R_1,\, R_2$, are dependent on the specific operator $\O^X$, and their values influence the extent of excited-state contamination in the matrix element. 
Smaller values of these constants or the presence of certain symmetries can lead to reduced excited-state contamination in the final result.

In the sum of ratios 
\begin{equation}
\begin{aligned}
    S(T,\tau_\mathrm{skip}) &= a\sum_{\tau = \tau_\mathrm{skip}}^{T-\tau_\mathrm{skip}} R(T,\tau) \\
    &= \mathcal{M}\left( T-\tau_\mathrm{skip} \right) + \text{excited states},
    \label{eq-definition-sum-ratios}
\end{aligned}
\end{equation}
excited-state contamination is exponentially suppressed with $T$ compared to $\nicefrac{T}{2}$ for the ratios themselves~\cite{capitani2010systematic,bulava2010b}.
Increasing $\tau_\mathrm{skip}$ reduces excited-state contamination.
Following the proportionality relation of the ratios and desired matrix element~\eqref{eq-matrix-element}, we can extract the latter by use of a finite difference. Neglecting excited states, one finds
\begin{equation}
    \mathcal{M} = \frac{S(T+\delta, \tau_\mathrm{skip}) - S(T, \tau_\mathrm{skip})}{\delta}. 
    \label{eq-definition-FD}
\end{equation}
Due to the available data we use a combination of $\nicefrac{\delta}{a} \in\{1,2,3\}$ depending on whether a neighbour $T+\delta$ is available.

The analysis is outlined as follows:

\textbf{Estimation of Ratios}: First, we calculate the ratios $R(T,\tau)$ and ratio sums $S(T,\tau_\mathrm{skip})$ for each operator. 
In the unpolarized (V) case we use
\begin{table}[H]\centering
\setlength{\tabcolsep}{1em}
\renewcommand{\arraystretch}{1.5}
\begin{tabular}{lcc}
   1. & $\tau_1^{(3)}$ &  $\frac{1}{2} \left[  \frac{ \O^V_{11} + \O^V_{22} + \O^V_{33} }{3} - \O^V_{44} \right]$, \\
   2. & $\tau_1^{(3)}$ &  $\frac{1}{\sqrt{2}} \left[  \O^V_{33} - \O^V_{44} \right]$, \\
   3. & $\tau_3^{(6)}$ &  $\frac{1}{\sqrt{2}} \left[  \O^V_{14} + \O^V_{41} \right]$,
\end{tabular}
\end{table}\noindent
further, in the polarized (A) case we use 
\begin{table}[H]\centering
\setlength{\tabcolsep}{1em}
\renewcommand{\arraystretch}{1.5}
\begin{tabular}{lcc}
    1. & $\tau_4^{(6)}$ &  $\frac{1}{\sqrt{2}} \left[  \O^A_{13} + \O^A_{31} \right]$, \\
    2. & $\tau_4^{(6)}$ &  $\frac{1}{\sqrt{2}} \left[  \O^A_{34} + \O^A_{43} \right]$,
\end{tabular}
\end{table}\noindent
and finally for the transversity (T) case we use 
\begin{table}[H]\centering
\setlength{\tabcolsep}{1em}
\renewcommand{\arraystretch}{1.5}
\begin{tabular}{lcc}
    1. & $\tau_1^{(8)}$ &  $\O^T_{211} - \O^T_{244}$, \\
    2. & $\tau_1^{(8)}$ &  $\O^T_{233} - \O^T_{244}$, \\
    3. & $\tau_2^{(8)}$ &  $\O^T_{124} - \O^T_{241}$, \\
    4. & $\tau_2^{(8)}$ &  $\O^T_{142} + \O^T_{421} - 2 \O^T_{214}$. \\
\end{tabular}
\end{table}\noindent
These have been carefully chosen to have nonzero kinematic factors, compare Equation~\eqref{eq-matrix-element}, and to be linearly independent.

\textbf{Matrix Element Extraction}: In the next step we extract matrix elements $\mathcal{M}$ using two different methods.
\textit{Method 1}: We extract the slope via finite differences at a specific source-sink separation $T=T'$, compare~\eqref{eq-definition-FD}.
\textit{Method 2}: We obtain the matrix element from a simultaneous (over all source-sink separations) and fully correlated fit to the 2-state form, equation~\eqref{eq-2-state}.
A matrix element obtained through either method is denoted as $\mathcal{M}\vert_{T',\mathfrak{m}}$, where $\mathfrak{m}$ represents the extraction method. 
For the second method the $T'$ index can be ignored.

From fitting the $\Cpt{\tau}$ we can obtain the ground-state energy $E_0$ which is used to calculate the kinematic factor.
After this, we calculate the unrenormalized moment as
\begin{equation}
    \mathfrak{X}_{\O^X,p,\mathfrak{m}}(T') = \frac{\mathcal{M}\vert_{T',\mathfrak{m}}}{\bar{u}_{N(p)} \Gamma^X_{\{ \alpha} \im p^{\phantom{X}}_{\mu \}} u_{N(p)}}.\label{eq-moment-step}
\end{equation}
To simplify the following equations, we define a compound index $j = \left(\O^X,p,\mathfrak{m}\right)$ that runs over
all operators and momenta with nonzero kinematic factors as well as the different methods to obtain the matrix element.

\textbf{Renormalization Factors}: We determine the renormalization factors in RI-(S)MOM and match them to $\bar{\mathrm{MS}}(2\,\si{GeV})$; for details see appendix~\ref{apx:npr}. 
This allows us to express the renormalized moment as $\mathfrak{X}_{j}^\mathrm{ren}(T') = Z_{\O^X} \cdot \mathfrak{X}_{j}(T')$.

\textbf{Moment of PDF}: To obtain the second moments of PDFs, we define the central value as the weighted average of the different results:
\begin{equation}
    \expval{x}^\mathrm{ren} = 
            \sum_{\tiny j,T' \geq T_\mathrm{plat}^{j}} \mathfrak{W}_{j}(T') \mathfrak{X}_{j}^\mathrm{ren}(T') 
        .\label{eq-final-moment}
\end{equation}
Here $T_\mathrm{plat}^{j}$ denotes the smallest source-sink separation such that $\mathfrak{X}_{j}(T')$ agree for all $T'\geq T_\mathrm{plat}^{j}$.
Naturally, the sum over $T'$ does not apply for the second method, where we fit the 2-state function, as there is no $T'$ to consider. 
The weights $\mathfrak{W}_{j}(T') \propto \nicefrac{1}{\sigma^2_{j}(T')}$ are normalised in such a way that weights associated to sum ratios sum to \nicefrac{1}{2} as do the weights for the 2-state fit. 
The used variances are estimated via bootstrap over $\mathfrak{X}_{j}^{}(T')$ and the errors of the renormalization constants are propagated.

\textbf{Systematic Error Estimation}: Finally, we estimate a systematic error, constraining the uncertainty from excited state contamination, by calculating the weighted standard deviation over the different results:
\begin{equation}
    \sigma_\mathrm{syst}^2 =         
        \sum_{\tiny j, T'\geq T_\mathrm{plat}^j} \mathfrak{W}_{j}(T') \left[ \mathfrak{X}_{j}^\mathrm{ren}(T') - \expval{x}^\mathrm{ren} \right]^2
        \label{eq-syst-error}.
\end{equation}
Again the sum over $T'$ is not applied for the 2-state fit.

\textbf{Relation to Quark Spin-Orbit Correlations}:
The longitudinal quark spin-orbit correlation $L^q_\ell S^q_\ell $ in the proton (where the subscript $\ell$ denotes alignment with the direction of motion of the proton) is related to the generalized transverse momentum-dependent parton distribution (GTMD) $G_{11}^{q} $ \cite{Lorce:2011kd} as in Equation~(\ref{lsgtmd}),
which in turn can be related to the generalized parton distributions (GPDs) $\widetilde{H}^{q} $, $H^q $, $E_T^q $ and $\widetilde{H}_T^q $ \cite{Lorce:2014mxa,PhysRevD.98.074022} as in Equation~(\ref{lsgpd}),
\begin{align}
2\, L^q_\ell S^q_\ell &= \int_{-1}^{1} \dd{x} \int \dd^{2} k_T \frac{k_T^2 }{m_N^2 } G_{11}^{q}
    \label{lsgtmd} \\
    &= \frac{1}{2} \int_{-1}^{1} \dd{x} x \widetilde{H}^{q}       - \frac{1}{2} \int_{-1}^{1} \dd{x} H^{q} \nonumber \\
       &+ \frac{m_q}{2 m_N} \int_{-1}^{1} \dd{x} (E_T^q + 2 \widetilde{H}_T^q ) \ ,
        \label{lsgpd}
\end{align}
where all distribution functions are quoted according to the nomenclature of \cite{Meissner:2009ww} and are taken in the forward limit; $k_T $ denotes the quark transverse momentum. $\widetilde{H}^{q} $ is the standard chiral-even helicity GPD and $H^q $ is the standard chiral-even unpolarized GPD; $E_T^q $ and $\widetilde{H}_T^q $ are chiral-odd GPDs. The longitudinal quark spin-orbit correlation has been evaluated according to Equation~(\ref{lsgtmd}) in Ref.~\cite{Engelhardt:2021kdo}; on the other hand, the results of the present work can be used complementarily to access the correlation via Equation~(\ref{lsgpd}), which can be viewed as the axial analogue of Ji's sum rule for orbital angular momentum: At the physical pion mass, the term proportional to $m_q /m_N$ is negligible. In the forward limit, $\int \dd{x} H^q $ corresponds to the number of valence quarks, i.e., unity in the isovector, $u-d$ quark case considered here. Therefore, to an excellent approximation, one has
\begin{equation}
2\, L^q_\ell S^q_\ell =
\frac{1}{2} \left(\expval{x}^\mathrm{ren}_A - 1 \right) \ ,
\label{eq-spin-orbital-from-moment}
\end{equation}
where $\int \dd{x} \, x \widetilde{H}^{q} =\expval{x}_{A} $ in the forward limit has been identified.
The results obtained in the following section will be used to quantify the longitudinal quark spin-orbit correlation and will also be confronted with the results of Ref.~\cite{Engelhardt:2021kdo}.  

\textbf{Simulation Parameters:}
We use a tree-level Symanzik-improved gauge action with 2+1 flavour tree-level improved Wilson Clover fermions coupling via 2-level HEX-smearing. 
Detailed information about the simulation setup can be found in references~\cite{durrLatticeQCDPhysical2011,durrLatticeQCDPhysical2011a,hasanNucleonAxialScalar2019}.
Key simulation parameters are summarized in Table~\ref{tab-Simulation-Details}.
Two ensembles, coarse and fine, at the physical pion mass are used. 
These ensembles correspond to lattice spacings of $\SI{0.1163(4)}{\femto\meter}$ and $\SI{0.0926(6)}{\femto\meter}$, respectively.
As described in~\cite{hasanNucleonAxialScalar2019}, the smearing is done using Wuppertal smearing~\cite{gusken1990study} -- $\tilde{q} \propto \left(1+\alpha H\right)^Nq$ with $H$ being the nearest-neighbor gauge-covariant hopping matrix -- at $\alpha=3$ and $N = 60,100$ for the coarse and fine ensemble, respectively. 
For each ensemble, two-point and three-point correlation functions are calculated. 
These calculations involve source-sink separations ranging from approximately $\SI{0.3}{\femto\meter}$ to $\SI{1.4}{\femto\meter}$ for the coarse ensemble and approximately $\SI{0.9}{\femto\meter}$ to $\SI{1.5}{\femto\meter}$ for the fine ensemble.
Furthermore, we consider two different momenta:
    $\vec{p} = (p_x,0,0)$ with $p_x = 0, -2[\nicefrac{2\pi}{L}]$ for the coarse ensemble, and
                          with $p_x = 0,-1 [\nicefrac{2\pi}{L}]$ for the fine ensemble.

\begin{table*}
\caption{
    Details of the used ensembles.
    The ensembles are at the physical pion mass, $m_{\pi} \approx m_{\pi}^{phys}$.
    A larger and a smaller lattice spacing, labelled as ``Coarse'' and ``Fine'' respectively, are available. 
    The ensembles were generated with a tree-level Symanzik-improved gauge action 
    with 2+1 flavour tree-level improved Wilson Clover fermions 
    coupled via 2-level HEX-smearing~\cite{durrLatticeQCDPhysical2011,durrLatticeQCDPhysical2011a,hasanNucleonAxialScalar2019}. 
    Furthermore, the available source-sink separations ($T$) and momenta ($p_x$) 
    which are used in the calculation of the ratios, Equation~\eqref{eq-ratio-definition}, are displayed.
}
\label{tab-Simulation-Details}
\begin{tabular}{l cccccccccc}
    \toprule
    Ensemble & Size & $\beta$ & $a[\SI{}{\femto\meter}]$ & $m_{\pi} [\SI{}{\MeV}]$ &  $m_\pi L$ & $\nicefrac{T}{a}$ & $p_x[\nicefrac{2\pi}{L}]$ & $\Ncfg$ \\
    \midrule
    Coarse & $\num{48}^4$ & \num{3.31} & \num{0.1163(4)} & \num{136(2)} & \num{3.9} & $3,4,5,6,7,8,10,12$ & $0,-2$ & \num{212} \\
    Fine   & $\num{64}^4$ & \num{3.5 } & \num{0.0926(6)} & \num{133(1)} & \num{4.0} & $10,13,16$          & $0,-1$ & \num{427} \\
    \bottomrule
\end{tabular}
\end{table*}

\section{Estimation of Matrix Elements}\label{subsec-Ratios}

\begin{figure*}
\resizebox{\textwidth}{!}{%
\includegraphics{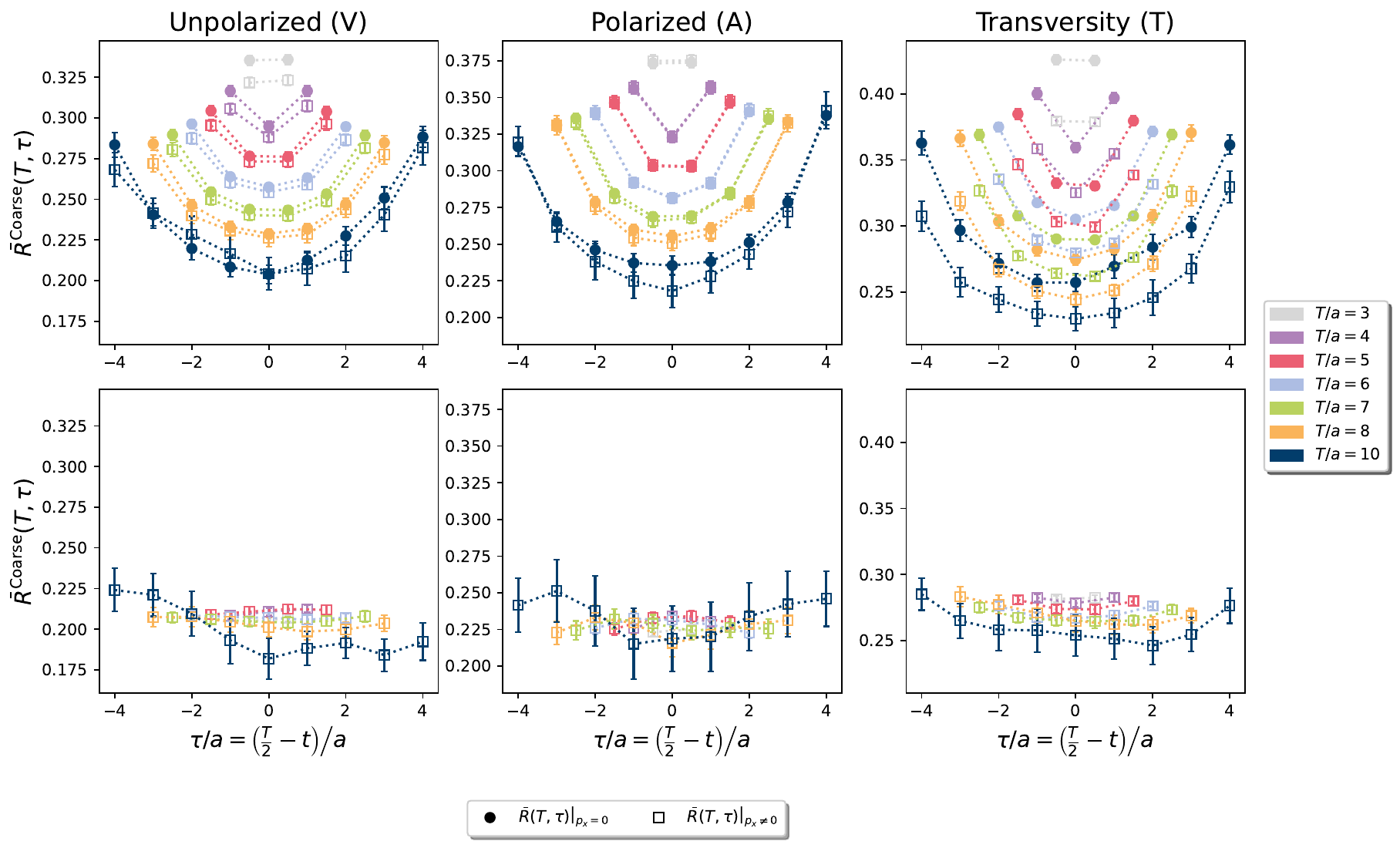}
}
\caption{
Ratios, cf.\@ Equation~\eqref{eq-ratio-definition}, for the coarse ensemble.
Various source-sink separations $T$ are represented by different colors, while the two momenta are distinguished using filled circles and unfilled squares. 
Each subplot corresponds to a different operator from the different channels organized by column.
For the Unpolarized (V) case we display operators 2.\@ and 3.\@; for the polarized (A) case we display operators 1.\@ and 2.\@; and for the transversity (T) case we display operators 3.\@ and 1.\@ for the upper and lower panel, respectively. 
}\label{Ratios} 
\end{figure*}

In Figure~\ref{Ratios}, we present results obtained from the coarse ensemble, using two different operators $\O^X$ per channel, as shown in the upper and lower rows.
Each column is dedicated to a particular channel: 
From top to bottom we display the operators 2.\@ and 3. (unpolarized), 1.\@ and 2.\@ (polarized), and 2.\@ and 3.\@ (transversity). 
Different source-sink separations are represented by various colours, while momenta are distinguished using filled circles for zero momentum and unfilled squares for finite momentum; this is kept consistent throughout all figures.
A plateau in these plots corresponds to the matrix element of the shown operator.
To simplify comparison we directly translate this to the bare moment, by multiplying with the kinematic factor 
$\bar{R}(T,\tau) = \nicefrac{1}{K} \cdot R(T,\tau)$. 
It is worth noting that we exclude the largest source-sink separation from these plots due to its substantial statistical uncertainty.

These operator choices are intentionally selected to illustrate the extreme variability of the excited-state contamination. 
While the upper row has a clearly visible cosh behavior -- as expected from the 2-state function~\eqref{eq-2-state} -- the lower row remains perfectly flat within statistics.  
Moreover, we observe that the convergence in source-sink separation is much faster for the lower row. 
For instance, in the lower row the plateau already converges after $\nicefrac{T}{a} = 3$ while the upper row requires $\nicefrac{T}{a} \geq 8$ in these particular examples.
This rapid convergence in the lower row is noteworthy, but it also comes with a drawback, which we observe in general, compare analysis summary plots in appendix~\ref{apx-results-per-operator} for the other operators: 
Operators that exhibit such flat behavior at small source-sink separations have a vanishing kinematic factor at zero momentum, making them computationally more challenging to handle.

\begin{figure*}
\resizebox{\textwidth}{!}{%
\includegraphics{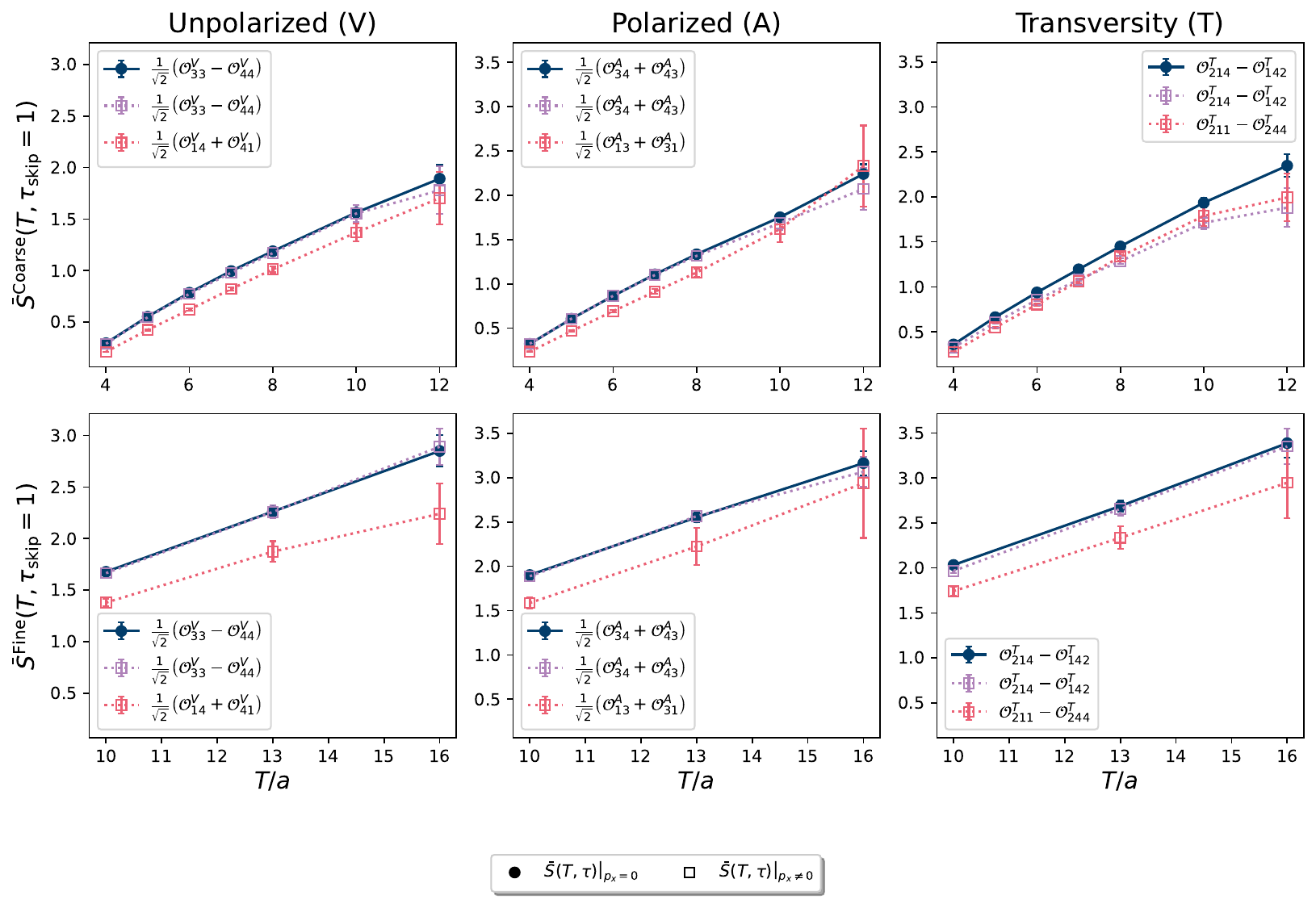}
}
\caption{
    Ratio sums $\bar{S}(T,\tau_\mathrm{skip})$ on the coarse and fine lattice,
    employing the same operators as in~\ref{Ratios}.
    Each $\bar{S}(T,\tau_\mathrm{skip})$ is plotted at fixed $\tau_\mathrm{skip} = 1$.
    As in Fig.~\ref{Ratios}, different momenta are displayed with hollow and filled markers.
}\label{Sum-Ratios}

\end{figure*}

In Figure~\ref{Sum-Ratios}, we present sum ratios, using the same operators as in Figure~\ref{Ratios} but put into one subplot.
The upper and lower row represent the coarse and fine ensemble, respectively.
The value of $\tau_\mathrm{skip} = 1$ is fixed as the slope of the summed ratios did not change for larger values.
The presence of excited-state contamination is subtly hinted at by the slight curvature observed in the data, 
although it is considerably less pronounced compared to the ratios.

\begin{figure*}
\resizebox{\textwidth}{!}{%
\includegraphics{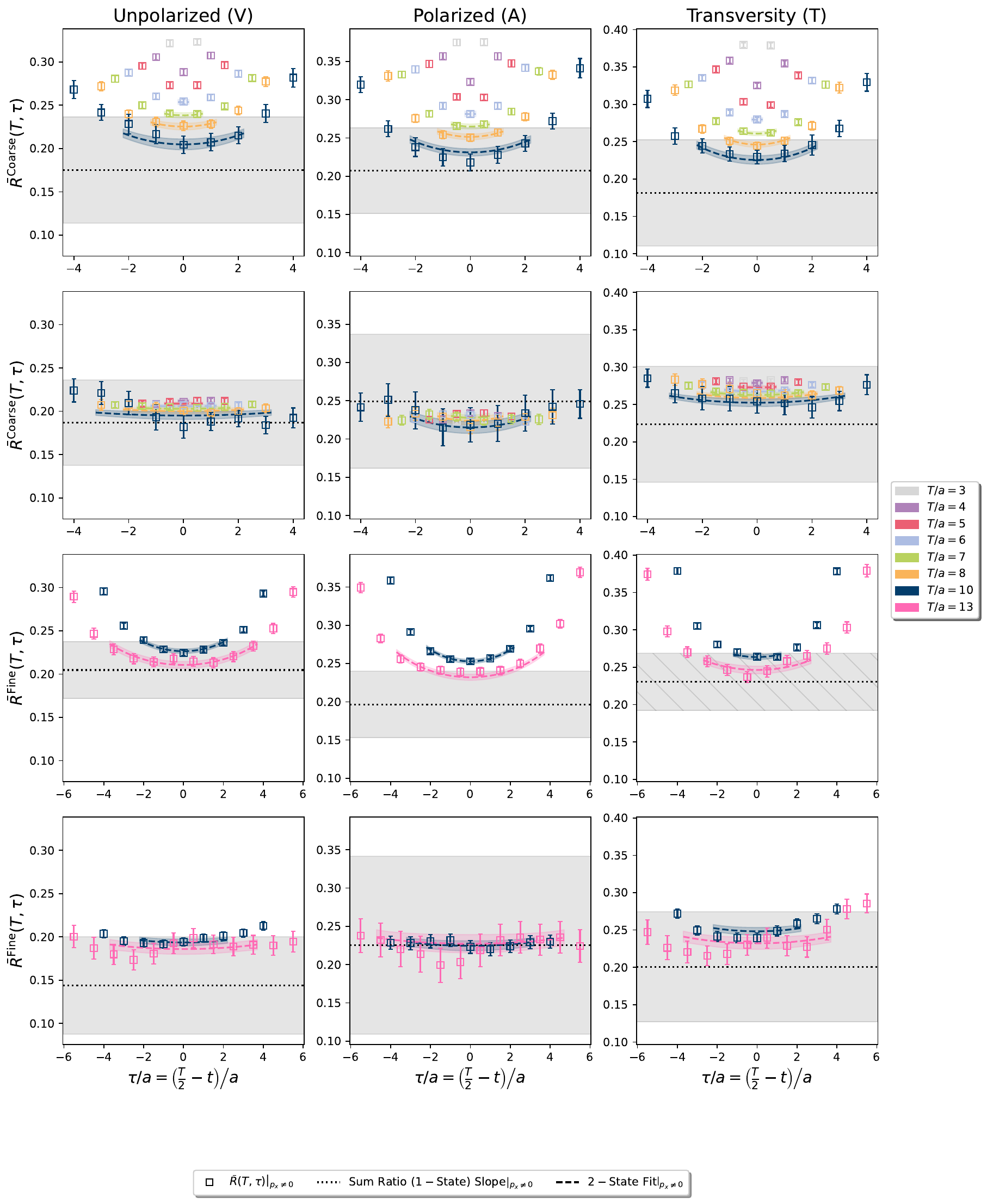}
}
\caption{
    Extraction result for the matrix elements, cf.\@ Equation~\eqref{eq-moment-step}, plotted on top of the original ratio data.
    The same operators and layout as in Figure~\ref{Ratios} are used. 
    Coarse and fine ensemble results are displayed in the first two and second two rows, respectively.
    Dot-dashed and dotted horizontal lines represent the average slope of the summed ratios divided by the kinematic factor.
    Solid and dashed lines represent the simultaneous and fully correlated central value fit to the ratios using the 2-state form~\eqref{eq-2-state}.
    Surrounding colored areas represent bootstrap uncertainties.
}\label{matrix-element}
\end{figure*}

In Figure~\ref{matrix-element}, we present the result of the matrix element extraction for the same operators as displayed in Figure~\ref{Ratios}.
Similar plots for all used operators can be found in appendix~\ref{apx-results-per-operator}.
We plot horizontal lines to represent the average (over $T'\geq T_\mathrm{plat}^j$) slope of the summed ratios, divided by the kinematic factor.
These slopes are extracted with the finite difference approach~\eqref{eq-definition-FD}.
As the matrix element is given by a plateau of the ratios, the expectation is that the plotted slope agrees at least with the central points $\tau \sim 0$ of large source-sink separations, which can be verified for all operators within uncertainty.
Again, those operators which are already flat match this expectation for more points and at smaller source sink separation.

Following the axolotl-like shape of the ratios, the solid, i.e.\@ zero momentum, and dashed, i.e.\@ finite momentum, lines indicate the central value 2-state fit result, using the form~\eqref{eq-2-state}.
The area around these indicate statistical uncertainty obtained via fitting on each bootstrap sample.
We use all data points that are covered by the best fit plot in a $(T,\tau)$-simultaneous fit.
This presents a fit interval in $\nicefrac{\tau}{a}$ which has been chosen by minimizing a $\nicefrac{\chi^2}{\mathrm{dof}}$.
The smaller source-sink separations for the coarse ensemble are excluded by this condition, as no points were left in the fit interval.

Considering all fits, values of $\nicefrac{\chi^2}{\mathrm{dof}}$ range from $\num{0.4}$ to $\num{2.7}$.
Correlations which go into these were estimated over the bootstrap samples of the included points and then kept fixed for the central value fit as well as the fits per bootstrap sample. 

Notably, the values of the matrix element obtained from summed ratios and 2-state fits always agree within statistics. 
The latter has reduced statistical uncertainty. 

\section{Moments of PDFs}\label{subsec-x}

\begin{figure*}
\resizebox{\textwidth}{!}{%
\includegraphics{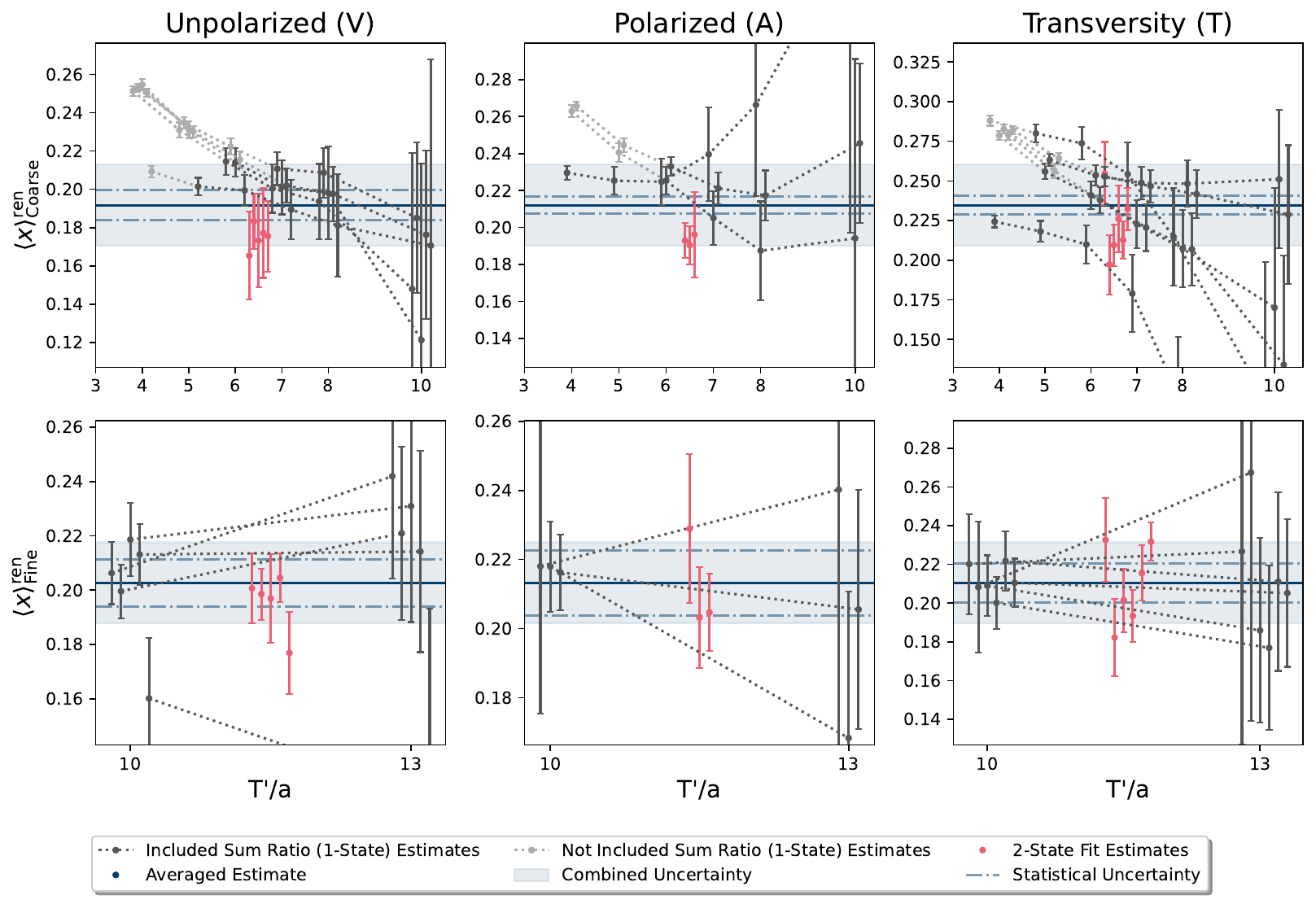}
}
\caption{
    Renormalized moments calculated from the summed ratio (1-state) (grey)~\eqref{eq-definition-sum-ratios}, and 2-state fit (red)~\eqref{eq-2-state}.
    The final average is displayed as a blue solid line while its statistical uncertainty is indicated via the blue dot-dashed line. 
    The blue band represents the statistical and systematic uncertainty, cf.\@ Equation~\eqref{eq-syst-error}, added in quadrature. 
    The light gray points are not included in the average as per the $T^j_\mathrm{plat}$ constraint, compare~\eqref{eq-final-moment}, to reduce excited-state effects.
    The ordinate limit is cut at $4 \sigma$ around the final average to increase resolution of the relevant points.
}\label{moments}
\end{figure*}

In Figure~\ref{moments}, we illustrate the results for the renormalized moments, which are extracted from the summed ratios (shown in grey, defined in Equation~\eqref{eq-definition-sum-ratios}) and the 2-state fits (displayed in red, as defined in Equation~\eqref{eq-2-state})\footnote{Summary plots, showing these results separated and labeled by their corresponding operators, momenta, methods and source-sink separations can be found in appendix~\ref{apx:summaryPlots}.
}.
The final average is denoted by the blue solid line, while its statistical uncertainty is indicated by the blue dot-dashed lines. The blue band represents the combined statistical and systematic uncertainty, as outlined in Equation~\eqref{eq-syst-error}, added in quadrature.
The light gray points are not included in the average, in accordance with the $T^j_\mathrm{plat}$ constraint.
To enhance the resolution of the relevant data points, the ordinate limit is truncated at $4\sigma$ and centered around the final average.
The numerical values of the final averages can be found in Table~\ref{tab-final-results}.

Comparing the two ensembles we find agreement within statistics indicating a flat continuum extrapolation.
With only two points a reliable extrapolation is not possible. 
The best we can do is to interpret the data points as Gaussian distributions, with mean equaling the central value and width given by the uncertainties added in quadrature, and perform a Bayesian fit. 
The relevant scale of discretization effects~\cite{gockeler2005lattice,capitani2001renormalisation} is $a \Lambda_\mathrm{QCD}$ resulting in a term proportional to $\alpha_s a \Lambda_\mathrm{QCD}$.
The operators themselves have tree-level quadratic discretization effects, 
resulting in the extrapolation
\begin{equation}
\begin{gathered}
    \expval{x}^\mathrm{ren}(a) = \expval{x}^\mathrm{ren}_\mathrm{cont} \cdot \\ 
    \Big(
        1 + m_1 \alpha_s a \Lambda_\mathrm{QCD} + m_2 \left(a \Lambda_\mathrm{QCD}\right)^2
    \Big)\label{eq:continuum-model}
\end{gathered}
\end{equation}
We use Gaussian priors for the coefficients, $p_{m_i} = \mathcal{N}(0,2)$ and no prior on the continuum value $\expval{x}^\mathrm{ren}_\mathrm{cont}$.
We approximate $\alpha_s \approx 0.3$ which is sufficient due the fact that the coefficients $m_i$ are mainly constrained by the prior. The continuum-extrapolated results are likewise given in Table~\ref{tab-final-results}.

\begin{table}[h]
    \caption{
        Final averages for the second moments of PDFs in the unpolarized, polarized and transversity channels, compare Figure~\ref{moments}.
        For the coarse and fine ensemble results,
        the central value is obtain as a weighted average over the different operators, momenta, and extraction methods, cf.\@ Equation~\eqref{eq-final-moment}.
        Further, the statistical uncertainty (first uncertainty) comes from a bootstrap over the original ensemble,
        while the systematic uncertainty (second uncertainty) is computed using the weighted standard deviation over the same set of results, cf.\@ Equation~\eqref{eq-syst-error}.
        We extrapolated the two points to the continuum limit using a Bayesian fit approach assuming them to be independent and Gaussian distributed with mean equaling the central value and standard deviation coming from the combined statistical and systematical uncertainty, compare Figure~\ref{continuum}.
    }\label{tab-final-results}
    \begin{tabular}{lp{6em}p{6em}}
    \toprule
     & Ensemble & $\expval{x}^{ren}$\\
     \midrule 
    Unpolarized (V) & Coarse    & $0.192(08)(20)$ \\
                    & Fine      & $0.203(09)(12)$ \\
                    & Continuum & $0.200(17)$     \\[1em]
    Polarized (A)   & Coarse    & $0.212(05)(21)$ \\
                    & Fine      & $0.213(09)(07)$ \\
                    & Continuum & $0.213(16)$     \\[1em]
    Transversity (T)& Coarse    & $0.235(06)(25)$ \\
                    & Fine      & $0.210(10)(18)$ \\
                    & Continuum & $0.219(21)$     \\[1em]
     \bottomrule
    \end{tabular}
\end{table}
\begin{figure*}
    \resizebox{\textwidth}{!}{
    \includegraphics{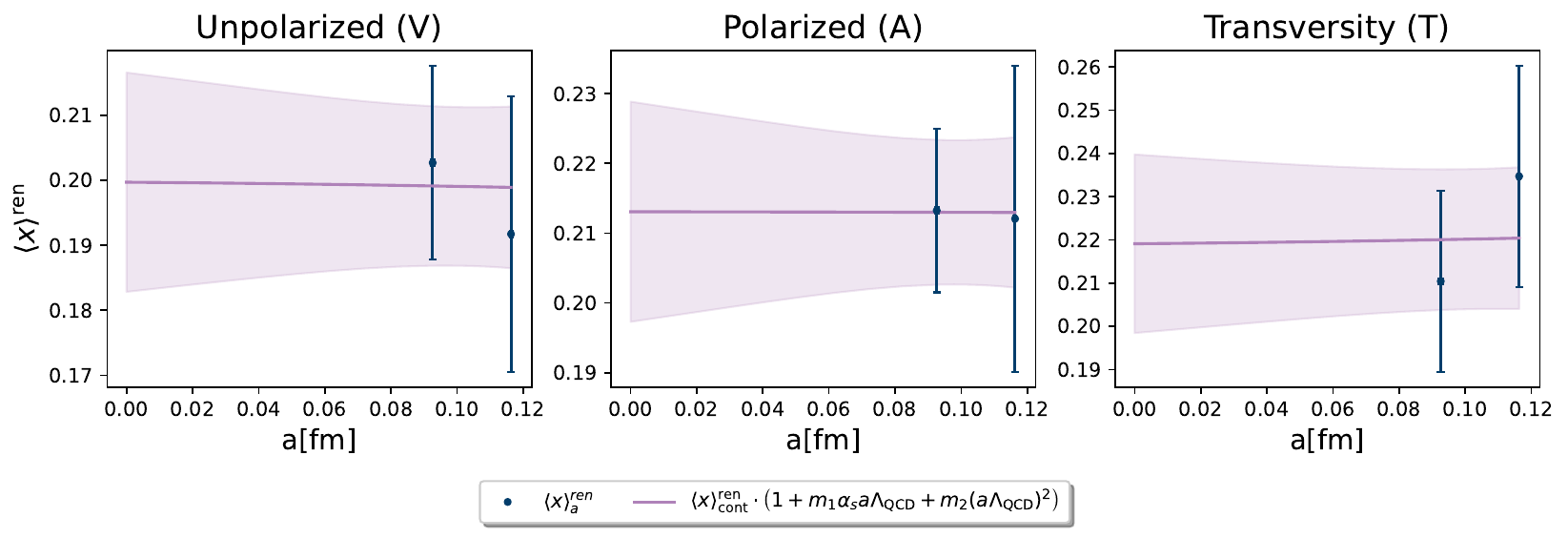}
    }\label{continuum}
    \caption{
        Continuum extrapolation using a Bayesian fit with the model described in Equation \eqref{eq:continuum-model}.
        The limited amount of data makes this extrapolation strongly dependent on the chosen priors for the coefficients $m_i$.
        Coming from a power counting these are expected to be of $\order{1}$, reflected in Gaussian priors of mean zero and variance 2.
        Resulting estimates are listed in Table~\ref{tab-final-results}.
    }
\end{figure*}

Our results are in good agreement, at the level of one to two standard deviations, with moments previously computed by other Lattice QCD collaborations~\cite{yang_proton_2018,harris_nucleon_2019,bali_nucleon_2019,alexandrou_moments_2020,alexandrou_complete_2020,mondal_moments_2020}.
Moreover, confronting our moments with phenomenological extractions, the comparison is quite favorable in the case of the axial moment, with Ref.~\cite{Blumlein:2010rn} giving $\langle x\rangle_{A} = 0.190 \pm 0.008$. On the other hand, in the unpolarized case, a certain tension between lattice and phenomenological results remains, with the recent determination in Ref.~\cite{li_extraction_2023}, $\langle x\rangle_{V} =0.143(5)$, differing from our result by about three standard deviations.

\section{Quark spin-orbit correlation}\label{subsec-LS}
With the results from Table~\ref{tab-final-results} we can calculate the longitudinal quark spin-orbit correlation in the proton according to Equation~(\ref{eq-spin-orbital-from-moment}). The obtained values can be found in Table~\ref{tab-spin-orbital-results}, along with the result obtained using the GTMD approach, Equation~(\ref{lsgtmd}), in Ref.~\cite{Engelhardt:2021kdo}. The results are in good agreement. As discussed in more detail in Ref.~\cite{Engelhardt:2021kdo}, the magnitude of this direct correlation between the spin and the orbital angular momentum of a quark significantly exceeds the correlation induced by the quark being embedded in a polarized proton environment. There is, therefore, a strong direct dynamical coupling between quark orbital angular momentum and spin, reminiscent of the $jj$ coupling scheme in atomic physics, rather than the Russell-Saunders coupling scheme.
\begin{table}[h]\centering
    \caption{
        Deduced isovector longitudinal quark spin-orbit correlation, estimated using the results for the polarized (A) moment shown in Table~\ref{tab-final-results} and relation~\eqref{eq-spin-orbital-from-moment}.
    }\label{tab-spin-orbital-results}
    \begin{tabular}{p{7em}c}
    \toprule
     Ensemble & $2 L^q_\ell S^q_\ell\left(\sigma_\mathrm{stat}\right)\left(\sigma_\mathrm{syst}\right)$ \\
    \midrule 
    Coarse    & $-0.394(02)(10)$  \\
    Fine      & $-0.393(05)(0)$  \\
    Continuum & $-0.393(08)$ \\
    GTMD$\vert_{a=0.114\mathrm{fm}}$~\cite{Engelhardt:2021kdo} & $-0.40 (2)$ \\
    \bottomrule
    \end{tabular}
\end{table}
 
\section{Summary}\label{sec-Summary}
In this study, we compute the second Mellin moment $\expval{x}$ of the unpolarized, polarized, and transversity parton distribution functions using two lattice QCD ensembles at the physical pion mass. 
Our approach involves extracting forward nucleon matrix elements at both zero and finite momentum, boosted in the x-direction.
Through the finite momentum data, we identify operators that exhibit remarkably small excited-state contamination.
Given the two ensembles a reliable continuum extrapolation is not accessible.
Regardless, we apply a Bayesian fit, accepting a strong dependence on the choice of priors, to provide a continuum estimate. 
The resulting values are in agreement with both individual ensembles: 
$\expval{x}_{u^+ - d^+}               = 0.200(17)$, 
$\expval{x}_{\Delta u^- - \Delta d^-} = 0.213(16)$, and
$\expval{x}_{\delta u^+ - \delta d^+} = 0.219(21)$.
Furthermore, we extract the isovector longitudinal quark spin-orbit correlation in the proton using the moment of the polarized PDF, 
$2L^q_\ell S^q_\ell = -0.393(08)$.
We find good agreement with earlier calculations based on GTMDs \cite{Engelhardt:2021kdo}.
 
\FloatBarrier
\begin{acknowledgments}
  We thank the Budapest-Marseille-Wuppertal Collaboration for making
  their configurations available to us and Nesreen Hasan for calculating
  the correlation functions analyzed here during the course of a different project.
  Calculations for this project
  were done using the Qlua software suite~\cite{Qlua}, and some of
  them made use of the QOPQDP adaptive multigrid
  solver~\cite{Babich:2010qb, QOPQDP}.  
  We gratefully acknowledge the computing time granted by the JARA Vergabegremium and provided on
  the JARA Partition part of the supercomputer JURECA~\cite{jureca} at
  Jülich Supercomputing Centre (JSC); computing time granted by the
  John von Neumann Institute for Computing (NIC) on the supercomputers
  JUQUEEN~\cite{juqueen}, JURECA, and JUWELS~\cite{juwels} at JSC; and
  computing time granted by the HLRS Steering Committee on Hazel Hen
  at the High Performance Computing Centre Stuttgart (HLRS).
  M.R.\@ was supported under the RWTH Exploratory Research Space (ERS) grant PF-JARA-SDS005 and MKW NRW under the funding code NW21-024-A.
  M.E.\@, S.L.\@, J.N.\@, and A.P.\@ are supported by the U.S. DOE Office of Science, Office of Nuclear Physics, through grants DE-FG02-96ER40965, DE-SC0016286, DE-SC-0011090, and DE-SC0023116, respectively.
  S.M.\@ is supported by the U.S. Department of Energy, Office of Science, Office of High Energy Physics under Award Number DE-SC0009913.
  S.S.\@ is supported by the National Science Foundation under CAREER Award PHY-1847893.
\end{acknowledgments}
 
\appendix
\twocolumngrid
\FloatBarrier
\section{Results Per Operator}\label{apx-results-per-operator}
\begin{figure*}
\resizebox{0.95\textwidth}{!}{%
    \includegraphics{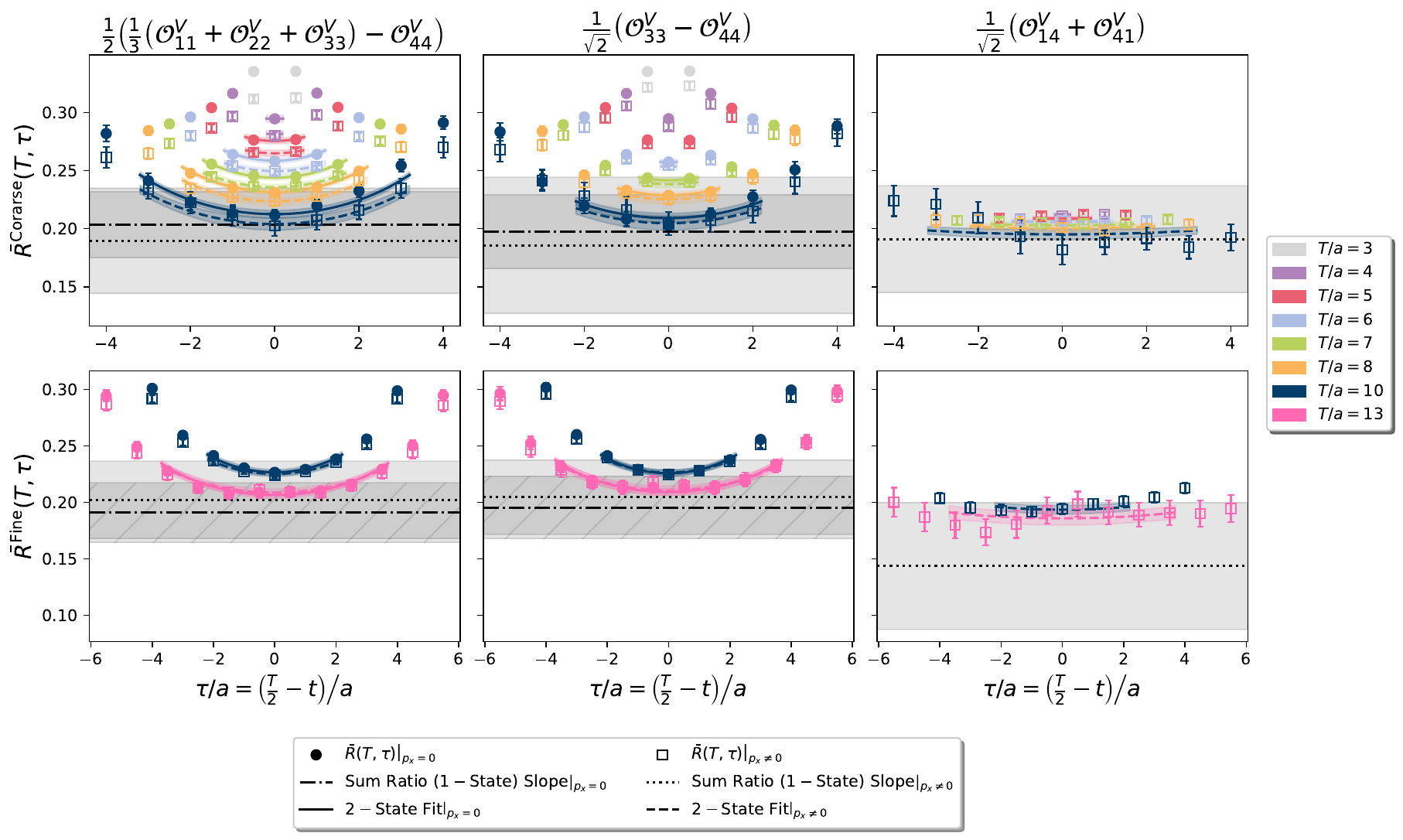}
}
\caption{
Analysis results of the ratios (points), slope of summed ratios (horizontal lines) and 2-state fit results (curves) for the operators 1., 2. and 3. corresponding to the unpolarized (vector) PDF.  }\label{fig-resperop-V}
\end{figure*}

\begin{figure*}
\resizebox{0.95\textwidth}{!}{%
    \includegraphics{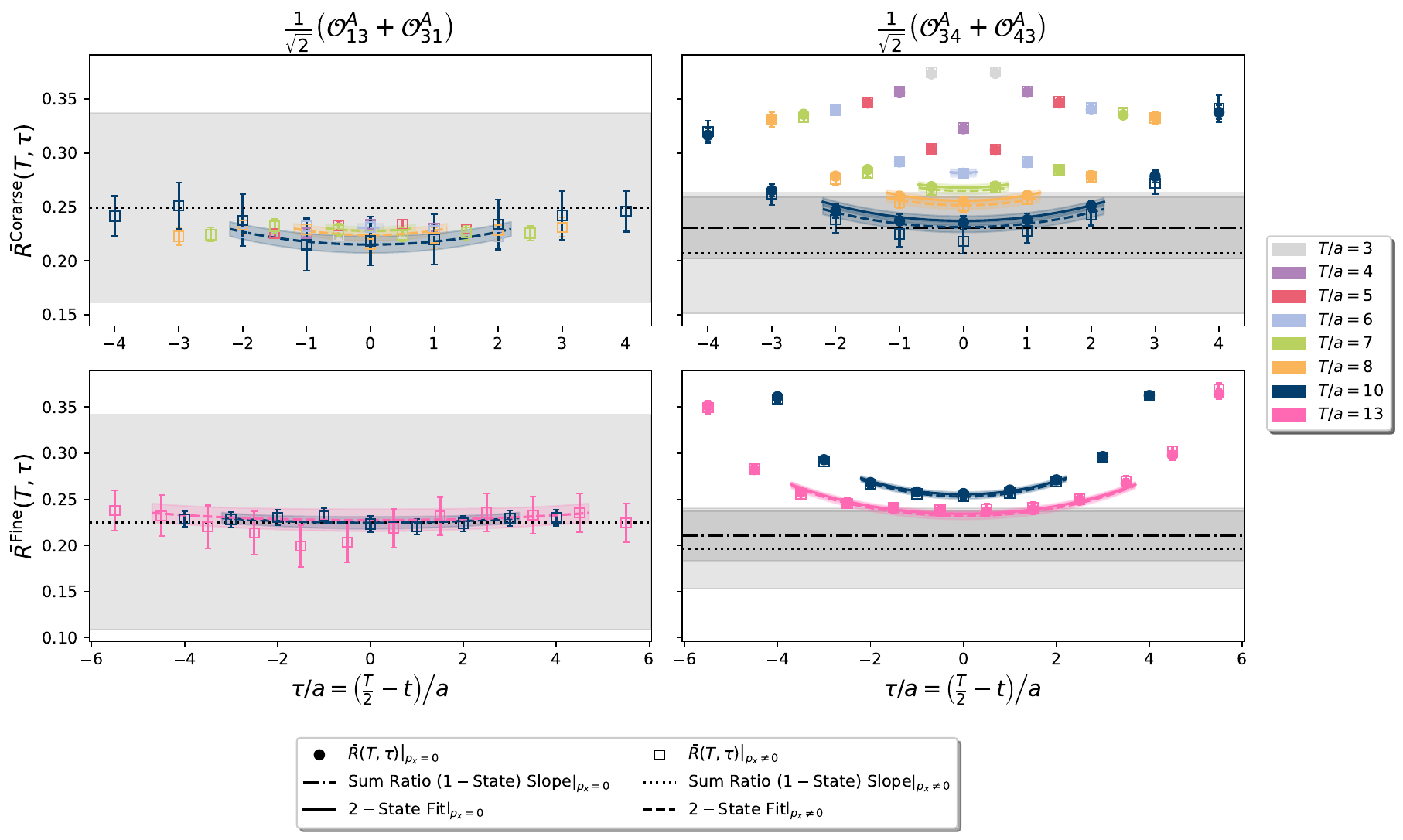}
}
\caption{
Analysis results of the ratios (points), slope of summed ratios (horizontal lines) and 2-state fit results (curves) for the operators 1. and 2. corresponding to the polarized (axial) PDF. 
}\label{fig-resperop-A}
\end{figure*}

\begin{figure*}
\resizebox{0.95\textwidth}{!}{%
    \includegraphics{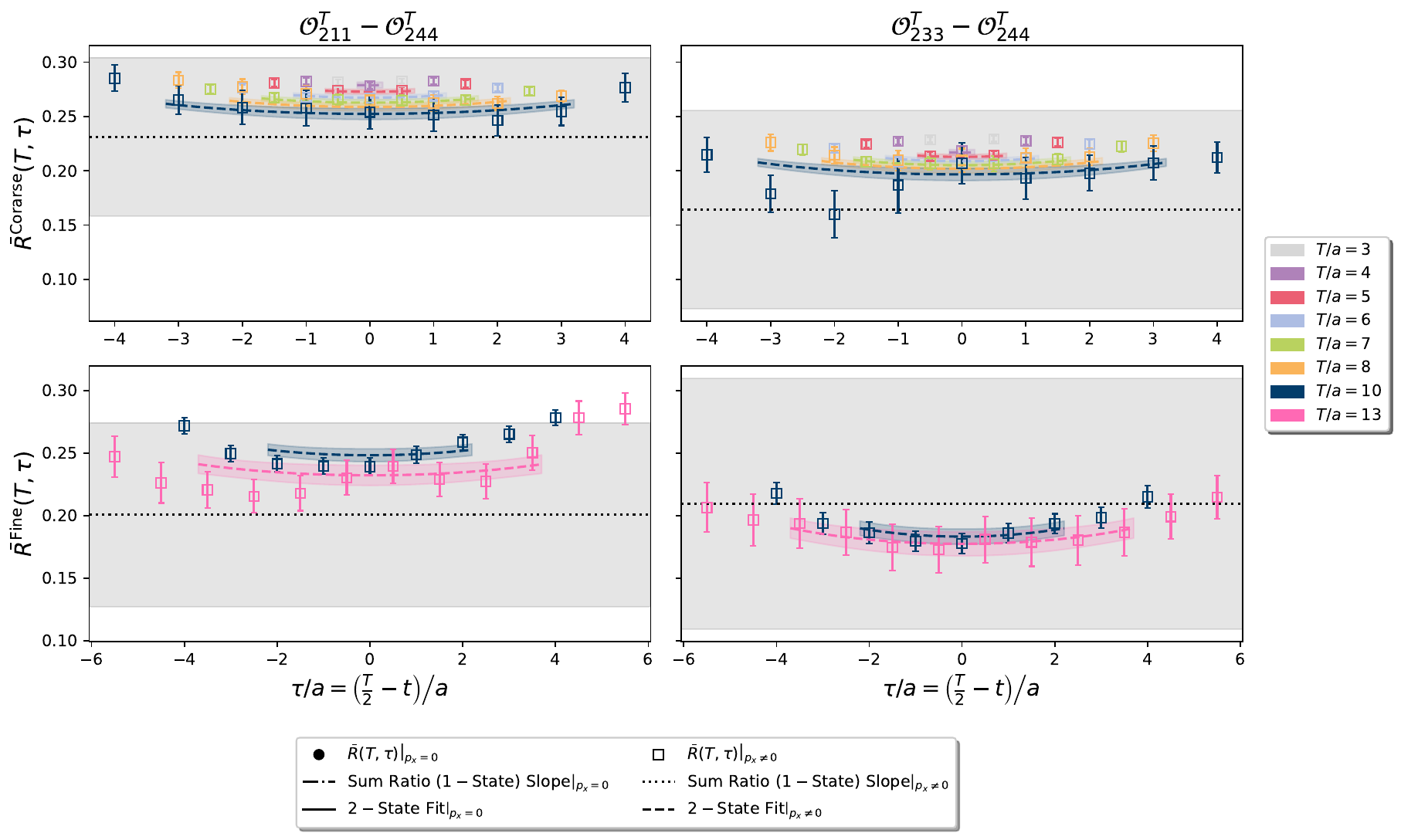}
}
\caption{
Analysis results of the ratios (points), slope of summed ratios (horizontal lines) and 2-state fit results (curves) for the operators 1. and 2. corresponding to the transversity (tensor) PDF. 
}\label{fig-resperop-T1}
\end{figure*}

\begin{figure*}
\resizebox{0.95\textwidth}{!}{%
    \includegraphics{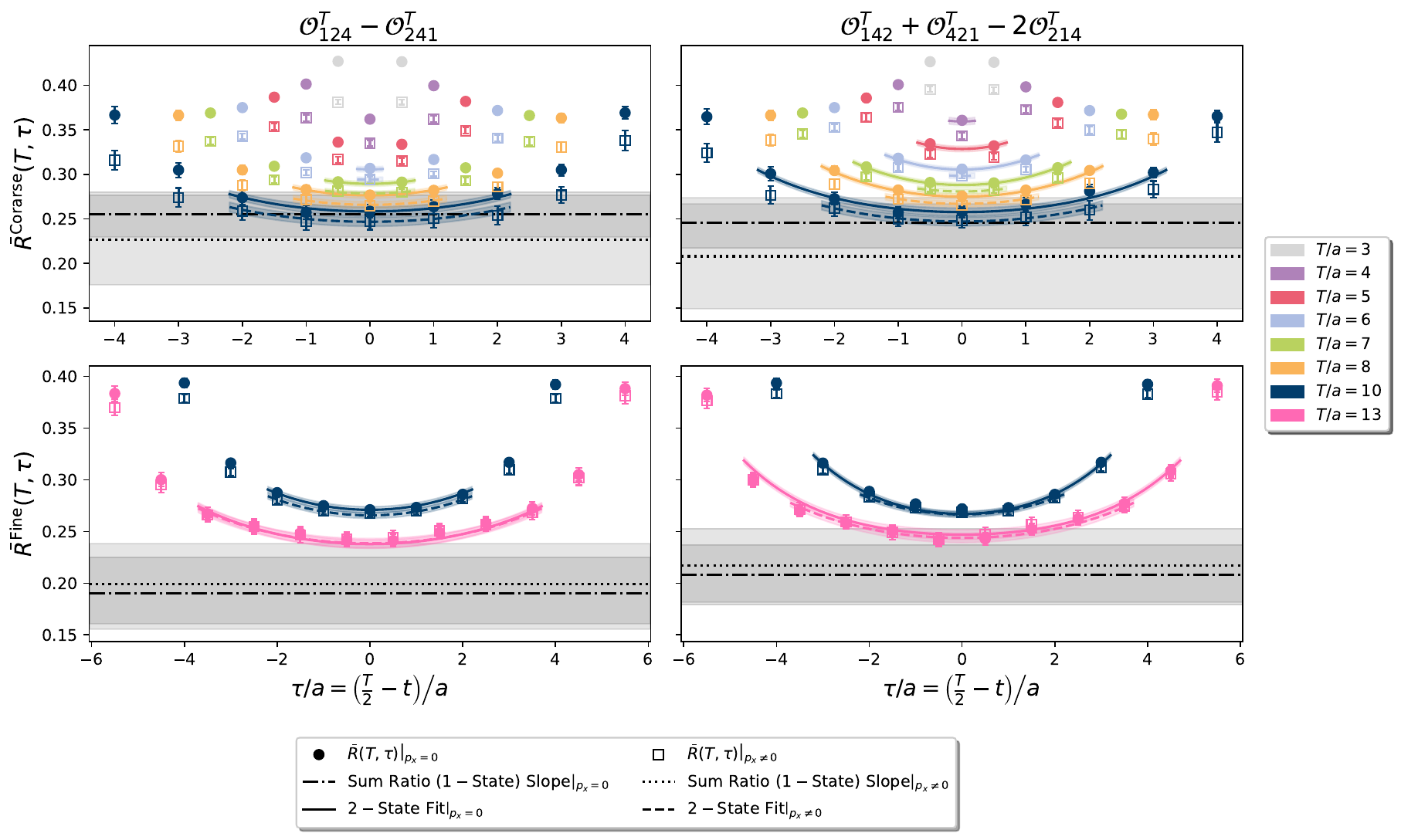}
}
\caption{
Analysis results of the ratios (points), slope of summed ratios (horizontal lines) and 2-state fit results (curves) for the operators 3. and 4. corresponding to the transversity (tensor) PDF. 
}\label{fig-resperop-T2}
\end{figure*}

In this appendix we show the analysis summary resolved per operator. 
As before, coarse and fine ensemble results are displayed in the first and second row, respectively.
Different colors represent different source sink separations and 
the horizontal dash-dotted, i.e.\@ zero momentum, and dotted, i.e.\@ finite momentum, lines represent the average (over $T'\geq T_\mathrm{plat}^j$) slope of the summed ratios, divided by the kinematic factor.
These slopes are extracted with the finite difference approach~\eqref{eq-definition-FD}.
The solid and dashed curves are the best-fit result of the 2-state fit to~\eqref{eq-2-state}, the surrounding band corresponds to the bootstrap uncertainty of the fit.
Figure~\ref{fig-resperop-V} displays the analysis of the operators corresponding to the unpolarized (vector) PDFs.
Figure~\ref{fig-resperop-A} displays the analysis of the operators corresponding to the polarized (axial) PDFs.
Figures~\ref{fig-resperop-T1} and~\ref{fig-resperop-T2} display the analysis of the operators corresponding to the transversity (tensor) PDFs.
As mentioned in Section~\ref{subsec-Ratios}, agreement of the slope of summed ratios with the plateau region expected around $\tau=0$ is given for all operators within one sigma. 
Corresponding best 2-state fit lines are in perfect agreement with the data points.

\section{Summary Plots}\label{apx:summaryPlots}

\begin{figure*}
    \centering
    \resizebox{\textwidth}{!}{\includegraphics{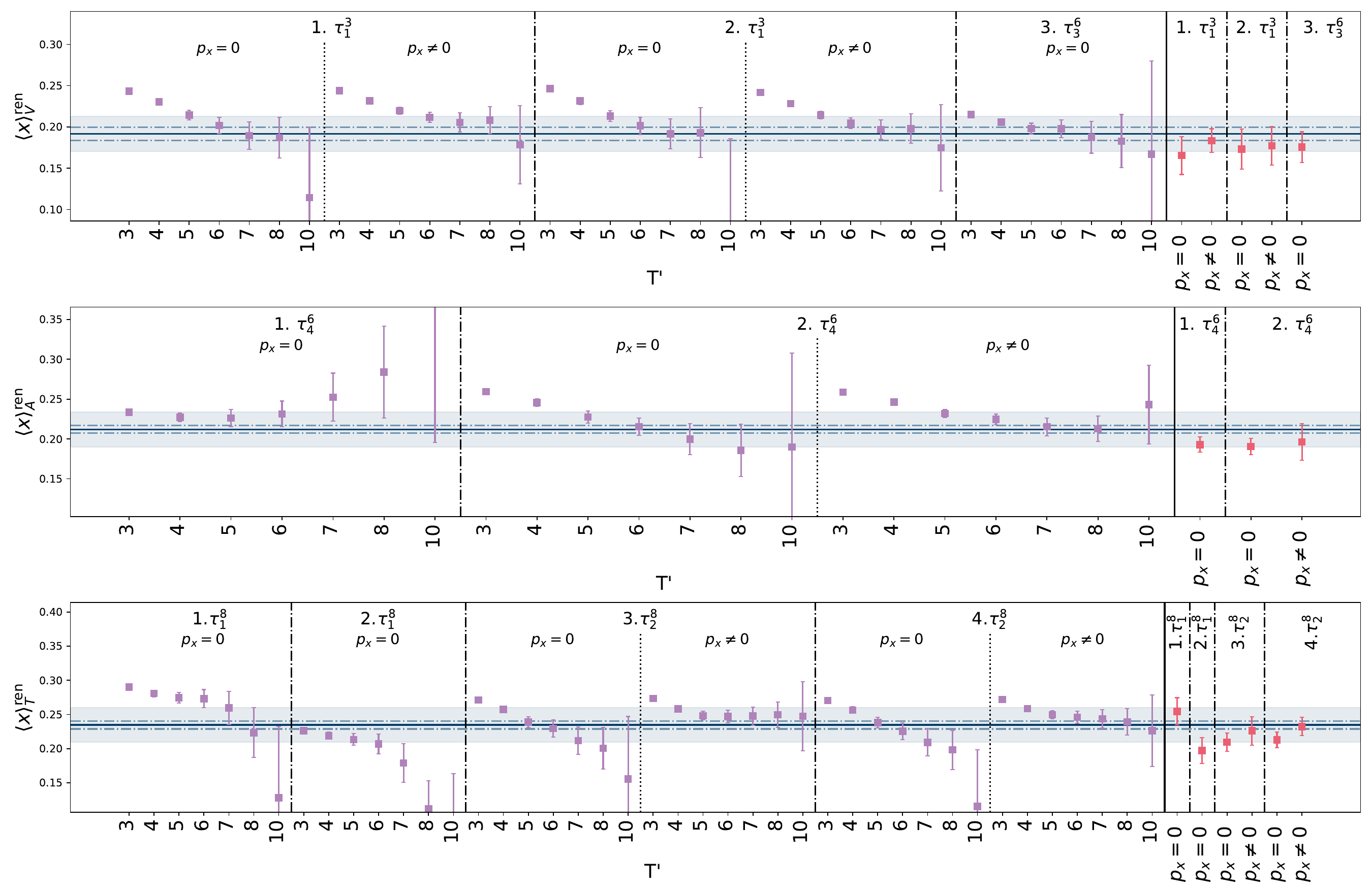}}
    \caption{
        Summary plot of the renormalized moment of PDF of the coarse ensemble resolved for each operator $\mathcal{O}^X$, represented by the corresponding ID and irrep,  and momentum $p_x$. 
        The purple points correspond to results obtained by the sum-ratio method evaluated at a source-sink separation $T'$.
        The red points are obtained using the two-state fit. 
        The blue solid line corresponds to the overall average as described by equation~\eqref{eq-final-moment} with the dashed line indicating statistical uncertainty and the blue area indicating statistical and systematic uncertainty added in quadrature.
    }
    \label{fig:x_summary_coarse}
\end{figure*}

\begin{figure*}
    \centering
    \resizebox{\textwidth}{!}{\includegraphics{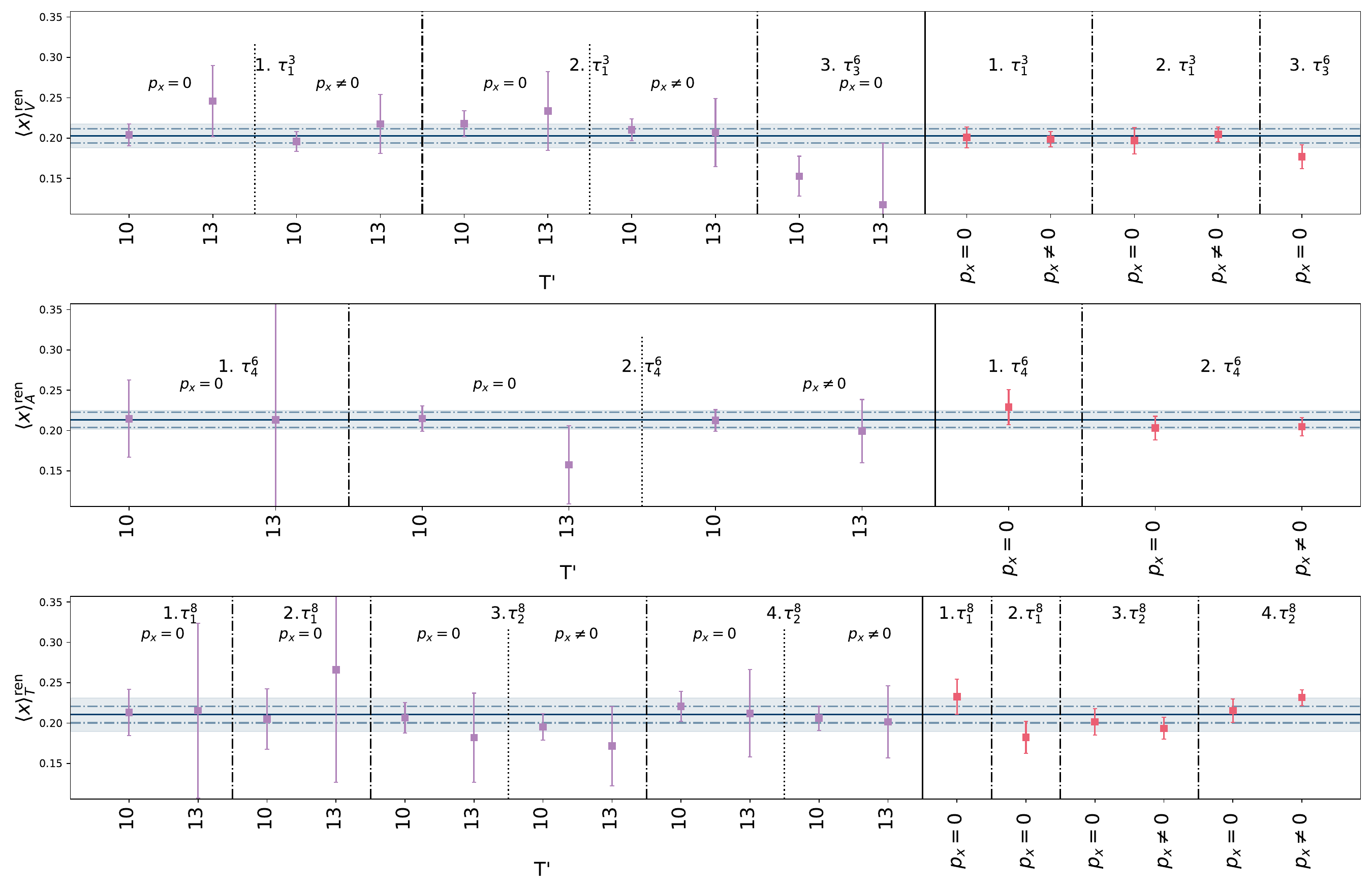}}
    \caption{
        Summary plot of the renormalized moment of PDF of the fine ensemble, similar to figure~\ref{fig:x_summary_coarse}
    }
    \label{fig:x_summary_fine}
\end{figure*}

We present summary plots of the moments for the coarse~\ref{fig:x_summary_coarse} and fine~\ref{fig:x_summary_fine} ensemble.
The three channels, unpolarized (V), polarized (A), and transversity (T) are shown in the columns.
Each result, i.e. the different operators and momenta, is displayed in the panels separated by the dotted and dashed lines.
The solid black line separates the sum-ratio method, points in purple, and the 2-state fit method, points in red. 
For the sum-ratio method the different $T'$ are spread across the abscissa.
As a point of reference, the average over the points, as described in equation~\eqref{eq-final-moment}, is shown by the horizontal blue line, with the statistical uncertainty shown by the dotted dashed line and the combined uncertainty by the blue band.
This corresponds to the blue line in figure~\ref{moments}.
A (strong) dependence on the source-sink separation can be seen in the sum-ratio related points. 

\FloatBarrier
\twocolumngrid
\section{Nonperturbative Renormalization}\label{apx:npr}

We determine renormalization factors for isovector vector, axial, and
tensor one-derivative twist-two operators using the nonperturbative
Rome-Southampton approach~\cite{Martinelli:1994ty}, in both
RI$'$-MOM~\cite{Martinelli:1994ty, Gockeler:1998ye} and RI-SMOM
schemes~\cite{Sturm:2009kb}, and convert and evolve to the \MSbar
scheme at scale 2~GeV using perturbation theory. We label these
renormalization factors $Z_{DV}^\rho$, $Z_{DA}^\rho$, and
$Z_{DT}^\rho$ for the one-derivative vector, axial, and tensor
operators, respectively, with $\rho$ denoting the irreducible
representation of the hypercubic group that takes on two possible
values in each case.

We largely follow our earlier work that used operators with no
derivatives~\cite{hasanNucleonAxialScalar2019}. Our primary data are
the Landau-gauge quark propagator,
\begin{equation}
  S(p) = \int d^4x\, e^{-ip\cdot x} \langle u(x) \bar u(0) \rangle,
\end{equation}
and the Landau-gauge Green's functions for operator $\O$,
\begin{equation}
  G_\O(p',p) = \int d^4x'\, d^4x\,
  e^{-ip'\cdot x'}e^{ip\cdot x} \langle u(x') \O(0) \bar u(x)\rangle.
\end{equation}
Here $\O$ is an isovector quark bilinear with one derivative,
yielding one Wick contraction: a connected diagram. We evaluate these
objects using four-dimensional volume plane wave sources, yielding an
effectively large sample size from the volume average. From these, we
construct our main objects, the amputated Green's functions,
\begin{equation}
  \Lambda_\O(p',p) = S^{-1}(p') G_\O(p',p) S^{-1}(p).
\end{equation}
Provided that $\O$ belongs to a definite irreducible
representation of the hypercubic group, these renormalize diagonally:
$\Lambda^R_\O =
(Z_\O/Z_\psi)\Lambda_\O$. To avoid determining
$Z_\psi$ directly, we will form ratios to determine
$Z_\O/Z_V$ and take $Z_V$ computed from pion three-point
functions in Ref.~\cite{hasanNucleonAxialScalar2019}.

\subsection{Conditions and matching}

The RI$'$-MOM scheme uses kinematics $p'=p$, whereas RI-SMOM uses
$p^2=(p')^2=q^2$ with $q=p'-p$. In both cases, the scale is defined as
$\mu^2=p^2$. For the vector current, we impose the conditions listed
in Ref.~\cite{hasanNucleonAxialScalar2019} on $\Lambda_{V_\mu}^R$ to
determine $Z_V/Z_\psi$.

For the one-derivative operators, we start with the continuum
decomposition of the amputated Green's function
$\Lambda_{\O_{\mu\nu\dots}}(p',p)$ into a sum of products of $O(4)$-invariant
functions $\Sigma^{(i)}_\O(p^2)$ and simple kinematic tensors
$\Lambda^{(i)}_{\O_{\mu\nu\dots}}(p',p)$. We then decompose
the operator and its kinematic tensors into irreducible
representations $\rho$, replacing $\mu\nu\dots$ with $\rho n$, where
$n$ ranges from 1 to the dimension of $\rho$. Tracing the amputated
Green's function with each of the tensors within each irrep, we get
\begin{multline}
  \sum_n \Tr\left[ \Lambda^{(i)}_{\O,\rho n}(p',p)
    \Lambda_{\O,\rho n}(p',p) \right] \\
  = M^{ij}_\rho(p',p) \Sigma^{(j)}_{\O,\rho}(p^2),
\end{multline}
where
\begin{equation}
  M^{ij}_\rho(p',p) = \sum_n \Tr\left[ \Lambda^{(i)}_{\O,\rho n}(p',p)
    \Lambda^{(j)}_{\O,\rho n}(p',p) \right]
\end{equation}
is a known kinematic matrix. Inverting $M$, we obtain the
$O(4)$-invariant functions computed within each irrep $\rho$,
$\Sigma^{(i)}_{\O,\rho}(p^2)$. Our choice of decomposition,
given below, is such that at tree level,
$\Sigma^{(i)}_\O(p^2)=\delta^{i1}$, and our renormalization
conditions will all be of the form
$\Sigma^{(1)}_{\O^R,\rho}(\mu^2)=1$. Basing this condition on
a $O(4)$-invariant function computed within each irrep ensures that
the ratio of renormalization factors for two different lattice irreps
of the same continuum operator is scale and scheme-invariant.

The one-derivative vector operator is
\begin{equation}
  \O^V_{\mu\nu} = \mathcal{S} \bar\psi \tau_3 \gamma_\mu\DLR_\nu \psi,
\end{equation}
where $\mathcal{S}$ takes the symmetric traceless part of the tensor:
\begin{equation}
  \mathcal{S} T_{\mu\nu} = \frac{1}{2}\left( T_{\mu\nu} + T_{\nu\mu} \right)
  - \frac{1}{4} \delta_{\mu\nu} T_{\alpha\alpha}.
\end{equation}
Our decomposition for the RI$'$-MOM scheme is a scaled version of the
one used by Gracey~\cite{Gracey:2003mr}:
\begin{align}
  \Lambda^{(1)}_{\O^V_{\mu\nu}}(p,p) &= \mathcal{S} \gamma_\mu p_\nu , \\
  \Lambda^{(2)}_{\O^V_{\mu\nu}}(p,p) &= \mathcal{S} \frac{p_\mu p_\nu}{p^2} \slashed{p},
\end{align}
where here and below we neglect tensors of opposite chirality. For
RI-SMOM, the derivative in the operators basis used by
Gracey~\cite{Gracey:2011zn} did not yield a definite $C$-symmetry,
unlike our operator containing $\DLR$. This allows us to use half as
many tensors as Gracey; see also \cite{Flynn:2014fea,
  Kniehl:2020nhw}. Defining $\bar p = (p'+p)/2$, our tensors and their
relation to Gracey's tensors $P^{W_2}_{(i)}$ are the following:
\begin{align}
  \Lambda^{S(1)}_{\O^V_{\mu\nu}}(p',p) &= \mathcal{S} \bar p_\mu \gamma_\nu
                       = \frac{1}{4} \left( P^{W_2}_{(2)} - P^{W_2}_{(1)} \right),\\
  \Lambda^{S(2)}_{\O^V_{\mu\nu}}(p',p) &=
     \mathcal{S} \frac{\bar p_\mu \bar p_\nu}{\bar p^2}\slashed{\bar p}
 = -\frac{1}{6} \sum_{i=3}^8 (-1)^i P^{W_2}_{(i)}, \\
  \Lambda^{S(3)}_{\O^V_{\mu\nu}}(p',p) &=
    \mathcal{S} \frac{\bar p_\mu q_\nu}{q^2}\slashed{q}
 = \frac{1}{2} \left( P^{W_2}_{(3)} - P^{W_2}_{(5)} + P^{W_2}_{(6)} - P^{W_2}_{(8)} \right),\\
  \Lambda^{S(4)}_{\O^V_{\mu\nu}}(p',p) &=
    \mathcal{S} \frac{\bar q_\mu q_\nu}{q^2}\slashed{\bar p}
 = \frac{1}{2} \left( \sum_{i=3}^5 P^{W_2}_{(i)} - \sum_{i=6}^8 P^{W_2}_{(i)} \right),\\
  \Lambda^{S(5)}_{\O^V_{\mu\nu}}(p',p) &=
    \mathcal{S} \frac{\bar p_\alpha q_\beta}{q^2} \gamma_{[\mu}\gamma_\alpha\gamma_{\beta]} \bar p_\nu
 = \frac{1}{2} \left( P^{W_2}_{(10)} - P^{W_2}_{(9)} \right),
\end{align}
where $\bar p^2=\frac{3}{4}\mu^2$ and the square brackets denote
antisymmetrization.  The one-derivative axial operator,
\begin{equation}
  \O^A_{\mu\nu} = \mathcal{S} \bar\psi \tau_3 \gamma_\mu \gamma_5 \DLR_\nu \psi,
\end{equation}
is related to the vector operator by chiral symmetry and its tensor
structures correspond to those of the vector operator, multiplied by
$\gamma_5$. We use the four-loop anomalous
dimension~\cite{Baikov:2006ai, Velizhanin:2011es,
  Baikov:2015tea}\footnote{Ref.~\cite{Velizhanin:2014fua} reports that
  Ref.~\cite{Velizhanin:2011es} contains a misprint.} and three-loop
matching to \MSbar~\cite{Gracey:2011zn, Kniehl:2020nhw}.

The one-derivative tensor operator is
\begin{equation}
  \O^T_{\mu\nu\rho} = \mathcal{S} \bar\psi \tau_3 \sigma_{\mu\nu} \DLR_\rho \psi,
\end{equation}
where the symmetrization and trace subtraction has the form~\cite{Gracey:2003mr}
\begin{multline}
  \mathcal{S} T_{\mu\nu\rho} =
  \frac{1}{2}\left( \tilde T_{\mu\nu\rho} + \tilde T_{\mu\rho\nu} \right)
    - \frac{1}{3} \delta_{\nu\rho} \tilde T_{\mu\alpha\alpha} \\
    + \frac{1}{6}\left( \delta_{\mu\nu} \tilde T_{\rho\alpha\alpha}
      + \delta_{\mu\rho} \tilde T_{\nu\alpha\alpha} \right)
\end{multline}
with
$\tilde T_{\mu\nu\rho} = \frac{1}{2}( T_{\mu\nu\rho} - T_{\nu\mu\rho}
)$.  Choosing to start by antisymmetrizing $\mu\nu$ leaves us with
only two tensor structures in the RI$'$-MOM scheme, compared with
Gracey's three:
\begin{align}
  \Lambda^{(1)}_{\O^T_{\mu\nu\rho}}(p,p) &= \mathcal{S} \sigma_{\mu\nu} p_\rho,\\
  \Lambda^{(2)}_{\O^T_{\mu\nu\rho}}(p,p) &= \mathcal{S}
                \frac{1}{p^2} \sigma_{\mu\alpha} p_\alpha p_\nu p_\rho.
\end{align}
For RI-SMOM, the supplementary data of Ref.~\cite{Braun:2016wnx} lists
30 structures. First antisymmetrizing $\mu\nu$ reduces this to 16 and
charge conjugation further reduces the number to 8. We write the first
six as
\begin{align}
\Lambda^{S(1)}_{\O^T_{\mu\nu\rho}}(p',p) &= \mathcal{S} \sigma_{\mu\nu} \bar p_\rho,\\
  \Lambda^{S(2)}_{\O^T_{\mu\nu\rho}}(p',p) &= \mathcal{S}
      \frac{1}{\bar p^2}\sigma_{\mu\alpha} \bar p_\alpha \bar p_\nu \bar p_\rho,\\
  \Lambda^{S(3)}_{\O^T_{\mu\nu\rho}}(p',p) &= \mathcal{S}
      \frac{1}{q^2}\sigma_{\mu\alpha} q_\alpha q_\nu \bar p_\rho,\\
  \Lambda^{S(4)}_{\O^T_{\mu\nu\rho}}(p',p) &= \mathcal{S}
      \frac{1}{q^2}\sigma_{\mu\alpha} q_\alpha \bar p_\nu q_\rho,\\
  \Lambda^{S(5)}_{\O^T_{\mu\nu\rho}}(p',p) &= \mathcal{S}
      \frac{1}{q^2}\sigma_{\mu\alpha} \bar p_\alpha q_\nu q_\rho,\\
  \Lambda^{S(6)}_{\O^T_{\mu\nu\rho}}(p',p) &= \mathcal{S}
     \frac{1}{q^2\bar p^2}\sigma_{\alpha\beta} \bar p_\alpha q_\beta
               \bar p_\mu q_\nu \bar p_\rho.
\end{align}
The last two tensors involve $\gamma_5$ or the identity, and they have
vanishing trace with each of the first six, so they can be
neglected. We use the three-loop anomalous
dimension~\cite{Gracey:2003mr}, the three-loop matching from
RI$'$-MOM~\cite{Gracey:2003mr}, and the two-loop matching from
RI-SMOM~\cite{Braun:2016wnx}.

\subsection{Calculation}

Our numerical setup follows Ref.~\cite{hasanNucleonAxialScalar2019},
extended to include both sets of momenta on both ensembles. We use
partially twisted boundary conditions, namely periodic in time for the
valence quarks. The plane wave sources are given momenta either along
the four-dimensional diagonal $p^{(\prime)}=\frac{2\pi}{L}(k,k,k,\pm k)$ or
along the two-dimensional diagonal
$p,p'\in\{\frac{2\pi}{L}(k,k,0,0),\frac{2\pi}{L}(k,0,k,0)\}$, with
$k=2,3,\dots,\frac{L}{4a}$. By contracting them in different
combinations, we get data for both RI$'$-MOM kinematics, $p'-p=0$, and
RI-SMOM kinematics, $p'-p=\frac{2\pi}{L}(0,0,0,\pm 2k)$ or
$\pm\frac{2\pi}{L}(0,k,-k,0)$. We used 54 gauge configurations from
each ensemble; however, on one configuration of the coarse ensemble
the valence twisted boundary condition yielded a near-singular Dirac
operator and the multigrid solver was unable to converge. We omitted
this configuration, using only 53 on the coarse ensemble.

\begin{figure*}
  \centering
  \label{fig:renorm}
  \includegraphics[width=0.49\textwidth]{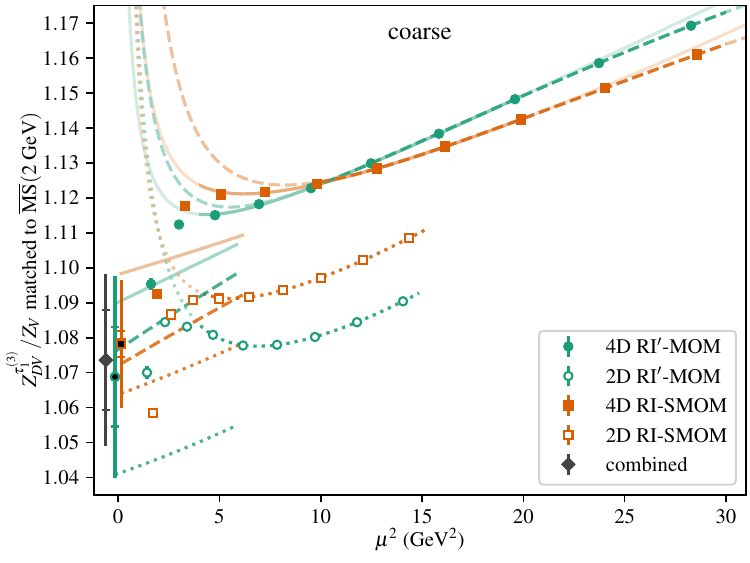}
  \includegraphics[width=0.49\textwidth]{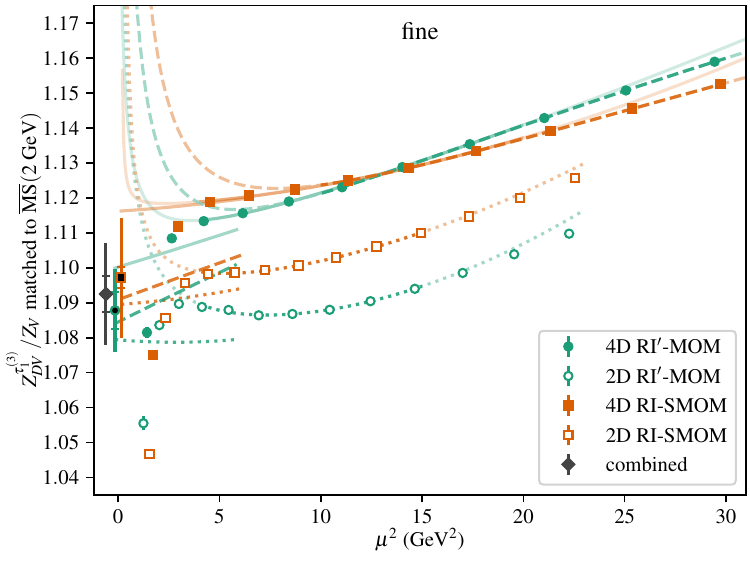}\\
  \includegraphics[width=0.49\textwidth]{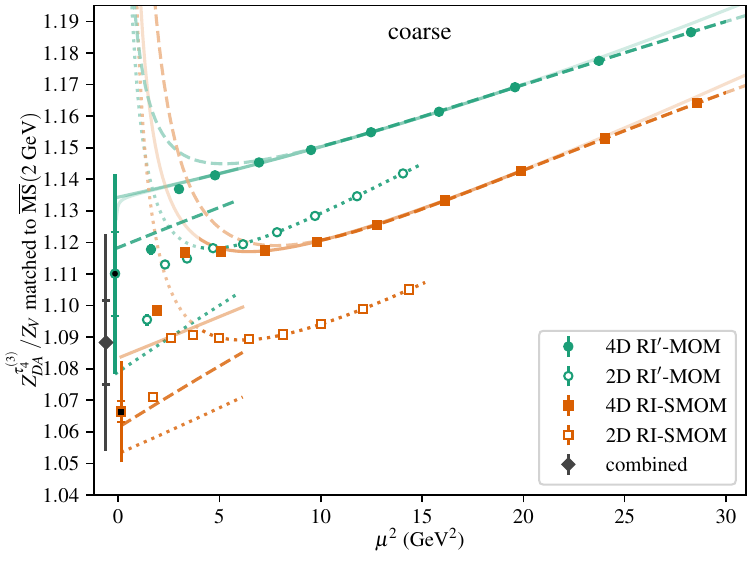}
  \includegraphics[width=0.49\textwidth]{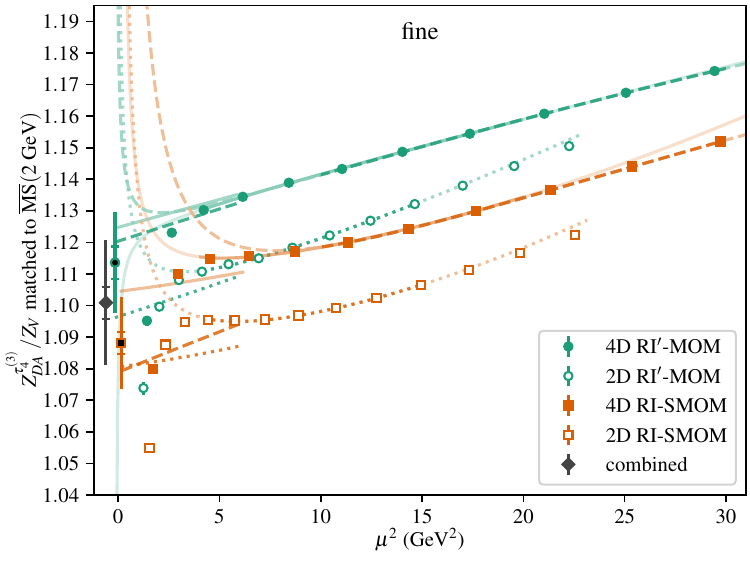}\\
  \includegraphics[width=0.49\textwidth]{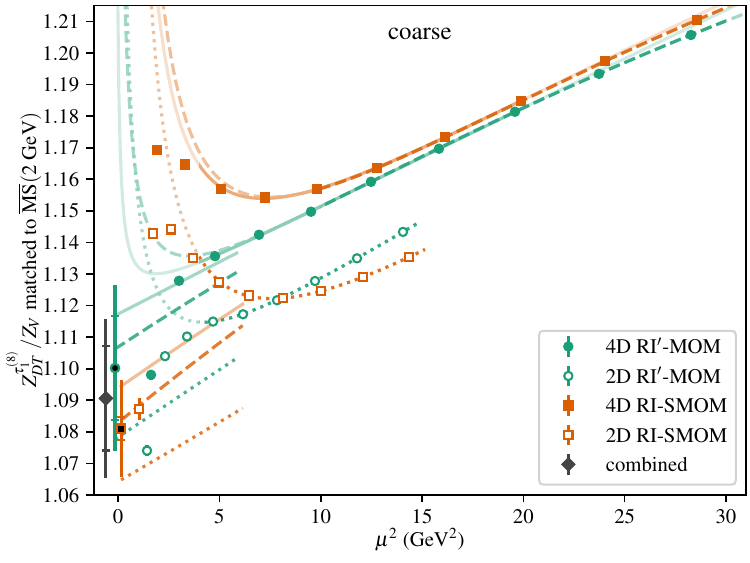}
  \includegraphics[width=0.49\textwidth]{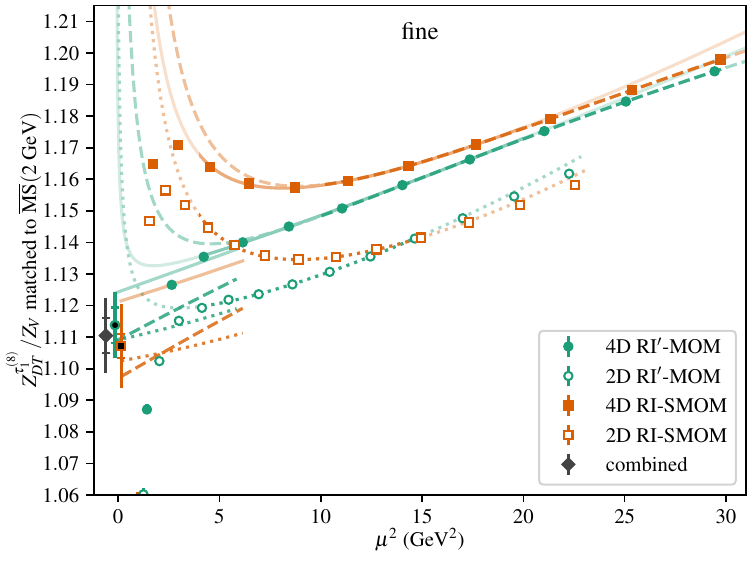}
    \caption{Ratios of renormalization factors
    $Z_{DV}^{\tau_1^{(3)}}/Z_V$, $Z_{DA}^{\tau_4^{(3)}}/Z_V$,
    $Z_{DT}^{\tau_1^{(8)}}/Z_V$ on the coarse (left) and fine (right)
    ensembles, determined using the RI$'$-MOM (green circles) and
    RI-SMOM (orange squares) intermediate schemes together with
    momenta along the four-dimensional diagonal (filled symbols) and
    two-dimensional diagonal (open symbols) and then matched to \MSbar
    at scale 2~GeV. For most points, the statistical uncertainty is
    smaller than the plotted symbol. The solid curves are fits to the
    4D data in the $\mu^2$ range from 4 to 20~GeV$^2$, the dashed
    curves in the range from 10 to 30~GeV$^2$, and the dotted curves
    are fits to the 2D data in the range from 4 to 15~GeV$^2$. The fit
    curves without the pole term are also plotted in the range
    $0 < \mu^2 < 6$~GeV$^2$. To reduce clutter, uncertainties on the
    fits are not shown. The symbols filled with black near $\mu^2=0$
    provide the final estimat for each intermediate scheme; their
    outer (without end cap) and inner (with end cap) error bars show
    the total and statistical uncertainties. The filled dark gray
    diamonds are the final estimates that combine both schemes.}
\end{figure*}

Perturbatively matching from RI$'$-MOM or RI-SMOM to the \MSbar scheme
and evolving to scale $\SI{2}{\GeV}$ does not eliminate dependence on
the initial scale $\mu$: there are remaining effects from lattice
artifacts, truncation of the perturbative series, and nonperturbative
contributions. To control these artifacts, we perform fits including
terms polynomial in $\mu^2$ and also a pole term, following
Ref.~\cite{Boucaud:2005rm}. These fits have the form
$A + B\mu^2 + C\mu^4 + D/\mu^2$, where the constant term $A$ serves as
our estimate of the renormalization factor ratio $Z_\O/Z_V$. Results
for this ratio, choosing one irrep for each of the three operator
types, are shown in Fig.~\ref{fig:renorm}. For each operator and
scheme, we perform three fits: using the 4D data with two different
fit ranges $\mu^2\in[4,20]$ and $[10,30]\text{ GeV}^2$ and using the
2D data with $\mu^2\in[4,15]\text{ GeV}^2$. In some cases
(particularly using the very precise RI-SMOM data), the fit quality is
very poor and thus we scale the statistical uncertainty by
$\sqrt{\chi^2/\text{dof}}$ whenever this is greater than
one. Following the same prescription as in
Ref.~\cite{hasanNucleonAxialScalar2019}, we combine the results first
within each scheme and then for our final result using both schemes,
estimating the systematic uncertainty (which is dominant) from the
scatter of results and conservatively taking the maximum statistical
uncertainty. In all cases, there is good agreement between the two
schemes.

\begin{figure*}
  \centering
  \label{fig:renorm_ratio}
  \includegraphics[width=0.49\textwidth]{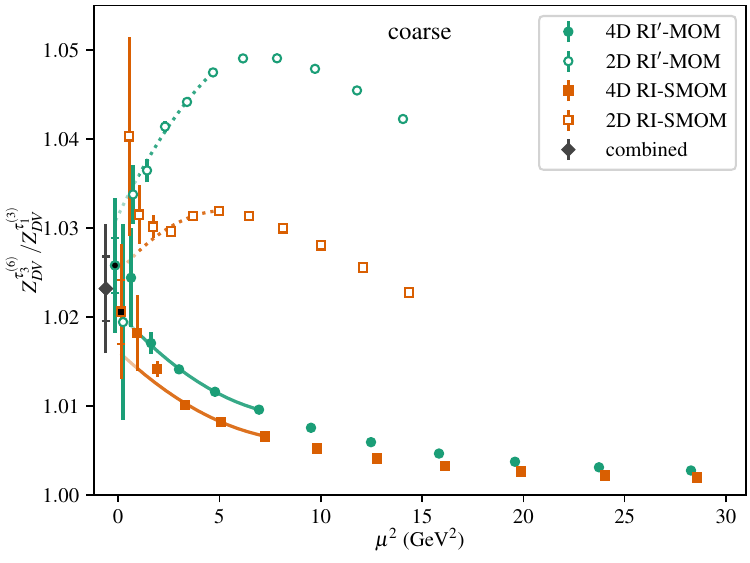}
  \includegraphics[width=0.49\textwidth]{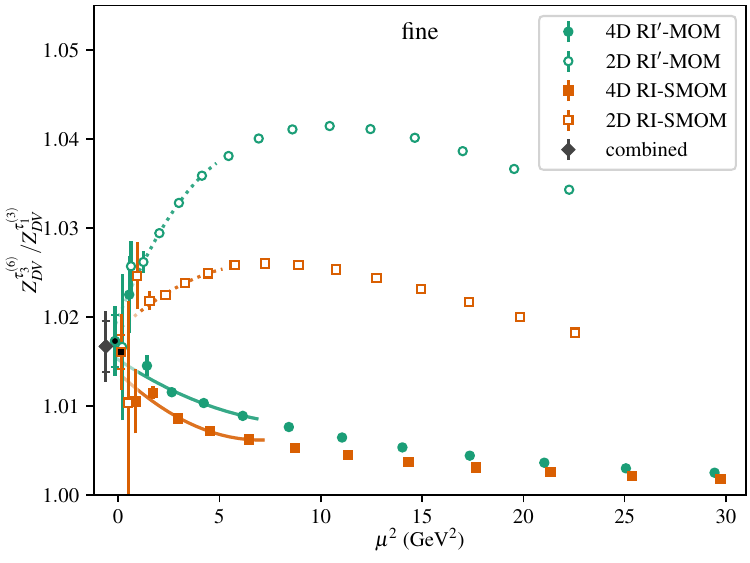}\\
  \includegraphics[width=0.49\textwidth]{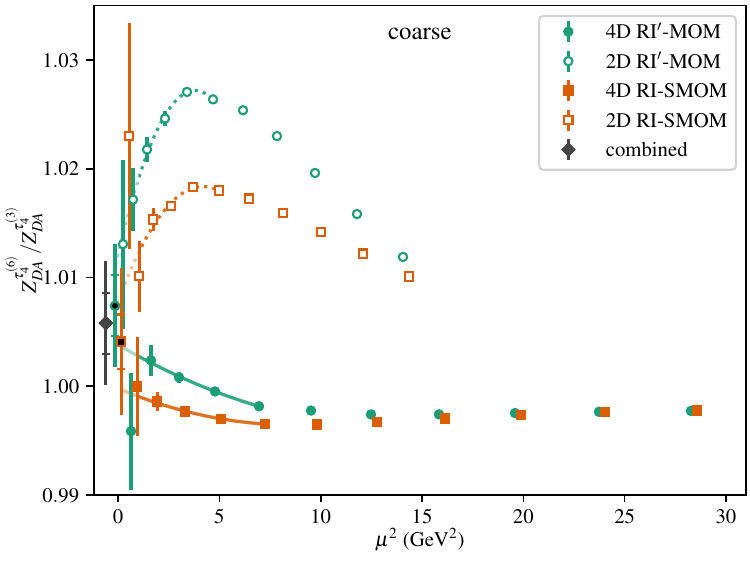}
  \includegraphics[width=0.49\textwidth]{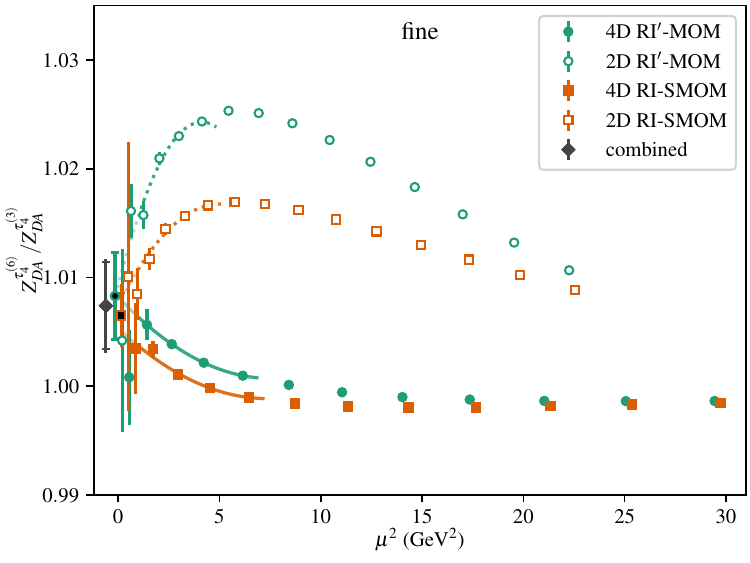}\\
  \includegraphics[width=0.49\textwidth]{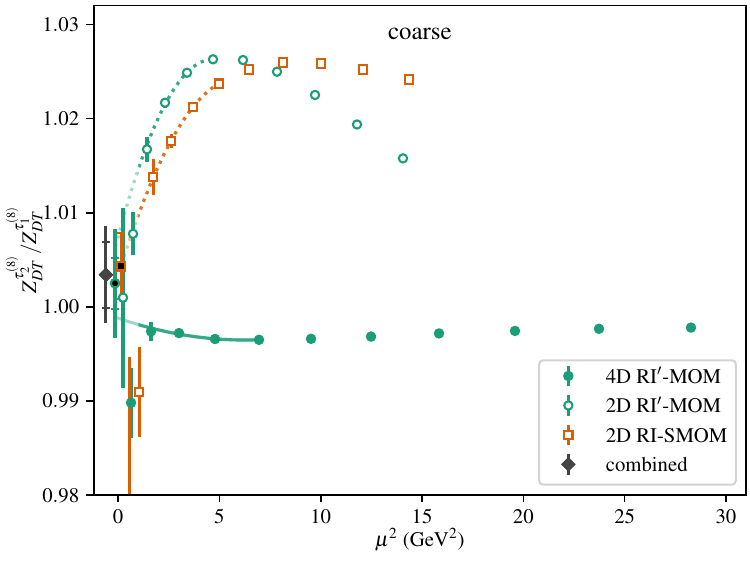}
  \includegraphics[width=0.49\textwidth]{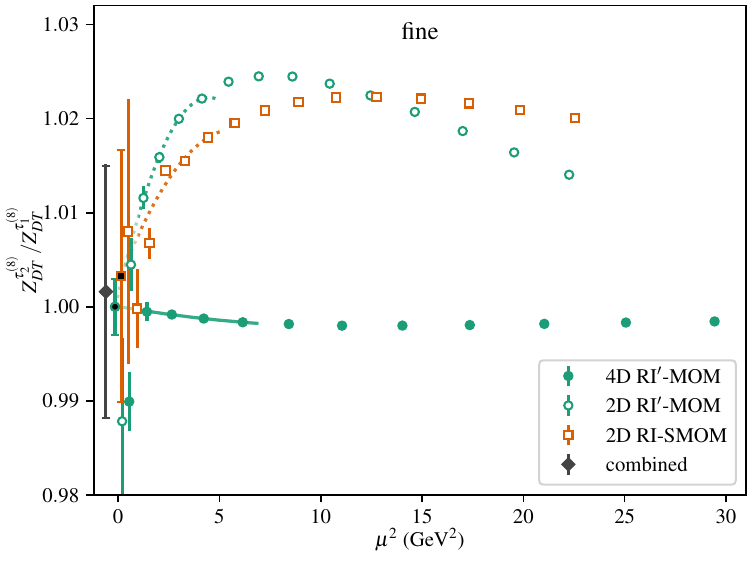}
  \caption{Scale and scheme-invariant ratios of renormalization
    factors $Z_{DV}^{\tau_3^{(6)}}/Z_{DV}^{\tau_1^{(3)}}$,
    $Z_{DA}^{\tau_4^{(6)}}/Z_{DA}^{\tau_4^{(3)}}$, and
    $Z_{DT}^{\tau_2^{(8)}}/Z_{DT}^{\tau_1^{(8)}}$, determined using
    the RI$'$-MOM (green circles) and RI-SMOM (orange squares)
    intermediate schemes together with momenta along the
    four-dimensional diagonal (filled symbols) and two-dimensional
    diagonal (open symbols) and then matched to \MSbar at scale
    2~GeV. For most points, the statistical uncertainty is smaller
    than the plotted symbol. The solid curves are fits to the 4D data
    in the $\mu^2$ range from 1 to 8~GeV$^2$ and the dotted curves are
    fits to the 2D data in the range from 1 to 5~GeV$^2$. To reduce
    clutter, uncertainties on the fits are not shown. The symbols
    filled with black near $\mu^2=0$ provide the final estimate for
    each intermediate scheme; their outer (without end cap) and inner
    (with end cap) error bars show the total and statistical
    uncertainties. The filled dark gray diamonds are the final
    estimates that combine both schemes.}
\end{figure*}

Renomalization-group-invariant ratios of renormalization factors in
different irreps $Z^{\rho'}_\O/Z^{\rho}_\O$ are shown in
Fig.~\ref{fig:renorm_ratio}. Note that it is not possible to isolate
the chosen $O(4)$-invariant function for the tensor operator in irrep
$\tau_2^{(8)}$ in the RI-SMOM scheme using the 4D kinematics. Because
in the continuum and infinite volume these ratios are independent of
$\mu^2$, we can fit using much lower momenta and only avoid the region
$\mu^2<\SI{1}{\GeV^2}$ due to finite-volume effects. We also omit the
pole term, i.e.\ set $D=0$. In all cases, we obtain results within a
few percent of unity.

\begin{table}
  \centering
  \label{tab:renorm}
  \caption{Vector current renormalization factor from
    Ref.~\cite{hasanNucleonAxialScalar2019} and ratios of
    renormalization factors computed in this work.}
  \begin{ruledtabular}
  \begin{tabular}{l ll}
    &coarse&fine\\
    \hline
$Z_V$                                     & 0.9094(36)       & 0.9438(1) \\
$Z_{DV}^{\tau_1^{(3)}}/Z_V$  \footnotemark[1] & 1.0736(142)(202) & 1.0925(52)(137)\\
$Z_{DV}^{\tau_3^{(6)}}/Z_{DV}^{\tau_1^{(3)}}$ & 1.0232(36)(63)   & 1.0167(29)(27)\\
$Z_{DA}^{\tau_4^{(3)}}/Z_V$  \footnotemark[1] & 1.0883(113)(316) & 1.1009(51)(192)\\
$Z_{DA}^{\tau_4^{(6)}}/Z_{DA}^{\tau_4^{(3)}}$ & 1.0058(28)(50)   & 1.0074(40)(16)\\
$Z_{DT}^{\tau_1^{(8)}}/Z_V$  \footnotemark[1] & 1.0906(165)(191) & 1.1105(56)(104)\\
$Z_{DT}^{\tau_2^{(8)}}/Z_{DT}^{\tau_1^{(8)}}$ & 1.0034(35)(38)   & 1.0016(134)(19)\\
  \end{tabular}
  \end{ruledtabular}
  \footnotetext[1]{\MSbar at 2 GeV}
\end{table}

Our final values for the ratios of renormalization factors are given
in Table~\ref{tab:renorm}.
 
\FloatBarrier

\makeatletter
\let\ORIbbl@fixname\bbl@fixname
\def\bbl@fixname#1{%
  \@ifundefined{languagealias@\expandafter\string#1}
    {\ORIbbl@fixname#1}
    {\edef\languagename{\@nameuse{languagealias@#1}}}%
}
\newcommand{\definelanguagealias}[2]{%
  \@namedef{languagealias@#1}{#2}%
}
\makeatother
\definelanguagealias{en}{english}

\end{document}